\documentclass[reprint, longbibliography]{revtex4-1}

\usepackage[utf8]{inputenc}

\usepackage[lining,semibold]{libertine} 
\usepackage{amsthm}
\usepackage[libertine, cmintegrals, bigdelims, vvarbb]{newtxmath}

\usepackage{amsmath}
\usepackage{amsfonts}
\usepackage{mathrsfs}
\usepackage{gensymb}
\usepackage{bbm}
\usepackage{dsfont}

\usepackage{kbordermatrix}
\renewcommand{\kbldelim}{(}
\renewcommand{\kbrdelim}{)}

\usepackage{blkarray}

\usepackage{chemformula}
\usepackage[caption=false]{subfig}

\usepackage{soul}
\usepackage{xcolor}

\theoremstyle{definition}

\usepackage{scalerel} 

\usepackage{comment} 

\definecolor{webgreen}{rgb}{0,.5,0}
\definecolor{webbrown}{rgb}{.6,0,0}
\definecolor{grigio}{rgb}{.85,.85,.85} 
\definecolor{RoyalBlue}{rgb}{0.0, 0.14, 0.4}
\definecolor{skyblue1}{rgb}{0.45,0.62,0.81}
\definecolor{skyblue2}{rgb}{0.2,0.39,0.64}
\definecolor{skyblue3}{rgb}{0.13,0.29,0.53}
\definecolor{scarlet1}{rgb}{0.93,0.16,0.16}
\definecolor{scarlet2}{rgb}{0.8,0,0}
\definecolor{scarlet3}{rgb}{0.64,0,0}

\definecolor{g}{gray}{0.50}

\usepackage{hyperref}
\hypersetup{%
    colorlinks=true, linktocpage=true, pdfstartpage=1, pdfstartview=FitV,%
    breaklinks=true, pdfpagemode=UseNone, pageanchor=true, pdfpagemode=UseOutlines,%
    plainpages=false, bookmarksnumbered, bookmarksopen=true, bookmarksopenlevel=1,%
    hypertexnames=true, pdfhighlight=/O,
    urlcolor=webbrown, linkcolor=RoyalBlue, citecolor=webgreen, 
    pdftitle={},%
    pdfauthor={Francesco Avanzini},%
    pdfsubject={},%
    pdfkeywords={},%
    pdfcreator={pdfLaTeX},%
    pdfproducer={LaTeX REVTeX}%
}

\newcommand{\eff}[1]{\overline{#1}}
\newcommand{\steady}[1]{\overline{#1}}
\newcommand{\osc}[1]{\tilde{#1}}

\newcommand{\dt}{\mathrm d_t}
\newcommand{\de}{\mathrm d}

\newcommand{\muteindex}{i}

\newcommand{\vect}{\boldsymbol\phi}
\newcommand{\vectel}{\phi}

\newcommand{\chemspecies}{\alpha}
\newcommand{\chemspeciesvec}{{\boldsymbol \chemspecies}}
\newcommand{\setchemspecies}{ Z}

\newcommand{\conc}{{\boldsymbol z}}
\newcommand{\setinternal}{X}
\newcommand{\setexchanged}{Y}
\newcommand{\intconc}{\boldsymbol x}
\newcommand{\exconc}{\boldsymbol y}
\newcommand{\terconc}{{\boldsymbol p}}
\newcommand{\ssterconc}{\steady{\boldsymbol p}}

\newcommand{\elrct}{\rho}
\newcommand{\setelrct}{ R}
\newcommand{\curr}{{\boldsymbol j}}
\newcommand{\currel}{j}

\newcommand{\excurr}{{\boldsymbol I}}
\newcommand{\excurrY}{{\boldsymbol I}^{\setexchanged}}
\newcommand{\excurrel}{{ I}}
\newcommand{\ssexcurr}{\steady{\boldsymbol I}}
\newcommand{\ssexcurrel}{{ I}}
\newcommand{\ssssexcurrel}{{\steady{ I}}}

\newcommand{\stcoeff}[1]{\boldsymbol \nu_{#1\elrct}}

\newcommand{\matS}{{\mathbb S}}
\newcommand{\colS}{{\boldsymbol{ S}}}
\newcommand{\elS}{{S}}
\newcommand{\matSX}{{\mathbb S}^{\setinternal}}
\newcommand{\matSY}{{\mathbb S}^{\setexchanged}}
\newcommand{\matSeff}{\hat{\matS}}
\newcommand{\elSeff}{\hat{S}}

\newcommand{\consquantity}{L}

\newcommand{\chempotential}{\mu}
\newcommand{\stchempotential}{\mu^\circ}
\newcommand{\epr}{\dot{\Sigma}}

\newcommand{\module}{{m}}

\newcommand{\modsetchemspecies}{\setchemspecies_\module}
\newcommand{\modsetinternal}{ Q_\module}
\newcommand{\modsetexchanged}{ P_\module}
\newcommand{\setterminal}{ P}
\newcommand{\modintconc}{{\boldsymbol q}_\module}
\newcommand{\modintconcpss}{\eff{\boldsymbol q}_\module}
\newcommand{\modintconcosc}{\osc{\boldsymbol q}_\module}
\newcommand{\modexconc}{{\boldsymbol p}_\module}

\newcommand{\modsetelrct}{\setelrct_\module}
\newcommand{\modcurr}{\curr_\module}
\newcommand{\modeffcurr}{\eff{\curr}_\module}
\newcommand{\modosccurr}{\osc{\curr}_\module}
\newcommand{\modexcurr}{\boldsymbol I_\module}
\newcommand{\modssexcurr}{{\boldsymbol I}_\module} 

\newcommand{\matSM}{\matS{}_\module}
\newcommand{\matSQM}{\matS{}^Q_\module}
\newcommand{\matSPM}{\matS{}^P_\module}
\newcommand{\effmatSPM}{\hat{\matS}{}^P_\module}
\newcommand{\matSeffQ}{\hat{\matS}{}^Q}
\newcommand{\matSeffP}{\hat{\matS}{}^P}

\newcommand{\modcycle}{{\boldsymbol c}}
\newcommand{\modcocycle}{{\boldsymbol v}}
\newcommand{\matCM}{{\mathbb C}_\module}
\newcommand{\modcycleindex}{{\gamma_\module}}
\newcommand{\modcocycleindex}{{\chi_\module}}
\newcommand{\modintcycleindex}{{\iota_\module}}
\newcommand{\modemcycleindex}{{\epsilon_\module}} 
\newcommand{\modcyclecurr}{\boldsymbol\psi}
\newcommand{\modcyclecurrel}{\psi}
\newcommand{\ssmodcyclecurrel}{\steady{\psi}}
\newcommand{\modcurrcycle}{{\boldsymbol{\modcyclecurr}}_\module}
\newcommand{\modeffcurrcycle}{\hat{\boldsymbol{\modcyclecurr}}_\module}
\newcommand{\cyclecurr}{\hat{\modcyclecurr}}
\newcommand{\sscyclecurr}{\steady{\modcyclecurr}}

\newcommand{\slowt}{\Delta t}
\newcommand{\period}{\tau}
\newcommand{\nosc}{n}

\newcommand{\lma}{a}
\newcommand{\lmb}{b}
\newcommand{\lmc}{c}
\newcommand{\lmd}{d}
\newcommand{\lme}{e}
\newcommand{\lmf}{f}
\newcommand{\lms}{s}
\newcommand{\lmge}{g_e}
\newcommand{\lmgf}{g_f}

\newcommand{\smodsscurrel}[1]{\steady{\currel}_{#1}}
\newcommand{\smodsscurr}[1]{\steady{\curr}_{#1}}

\newcommand{\smatSM}[1]{\matS{}_{#1}}
\newcommand{\smatSQM}[1]{\matS{}^Q_{#1}}
\newcommand{\smatSPM}[1]{\matS{}^P_{#1}}

\newcommand{\smodcyclecurrel}[1]{\modcyclecurrel_{#1}}

\newcommand{\freerct}[1]{\Delta_{#1} G}

\newcommand{\vfreerct}{\boldsymbol{\Delta}_{r} \boldsymbol{G}}

\newcommand{\cycle}{{\boldsymbol c}}
\newcommand{\cyclect}{\hat{\boldsymbol c}}
\newcommand{\inter}{{\iota}}
\newcommand{\emer}{{\epsilon}}

\newcommand{\nterminalm}{|P_\module|}
\newcommand{\nemerm}{| \emer_\module|}
\newcommand{\nbrokenm}{| \lambda_\module|}

\makeatletter
\def\maketag@@@#1{\hbox{\m@th\normalfont\normalsize#1}}
\makeatother

\DeclareMathAlphabet{\mathpzc}{OT1}{pzc}{m}{it}

\begin{document}

\title{Circuit Theory for Chemical Reaction Networks}
\newcommand\unilu{\affiliation{Complex Systems and Statistical Mechanics, Department of Physics and Materials Science, University of Luxembourg, L-1511 Luxembourg City, Luxembourg}}
\author{Francesco Avanzini}
\email{francesco.avanzini@uni.lu}
\unilu
\author{Nahuel Freitas}
\email{nahuel.freitas@uni.lu}
\unilu
\author{Massimiliano Esposito}
\email{massimiliano.esposito@uni.lu}
\unilu

\date{\today}

\begin{abstract}
We lay the foundation of a circuit theory for chemical reaction networks.
Chemical reactions are grouped into chemical modules solely characterized by their current-concentration characteristic, as electrical devices by their current-voltage (I-V) curve in electronic circuit theory.
This, combined with the chemical analog of Kirchhoff's current and voltage laws, provides a powerful tool to predict reaction currents and dissipation across complex chemical networks. 
The theory can serve to build accurate reduced models of complex networks as well as to design networks performing desired tasks. 
\end{abstract}

\maketitle



\section{Introduction}

Chemical reaction networks (CRNs) are ubiquitous in nature and can easily reach high levels of complexity.
Combustion~\cite{Gardiner1984}, atmospheric chemistry~\cite{Warneck1999,Wayne2006}, geochemistry~\cite{McSween2003}, biochemistry~\cite{voet2010}, biogeochemistry \cite{schlesinger2013,smith2016}, ecology~\cite{garvey2017}, provide some examples.  
The complexity of many of these networks arises from their large size and complex topology (encoded in the stoichiometric matrix), from the non-linearity of chemical kinetics, and from the fact that they do not operate in closed vessels.
They continuously exchange energy and matter with their surroundings thus maintaining chemical reactions out of equilibrium \cite{Hermans2017,Esposito2020}.
Their detailed characterization would require knowing the currents through all the reactions which, for elementary reactions satisfying mass-action kinetics, implies the knowledge of the reaction rate of every reaction and of the concentrations of all the species.
Naturally, such knowledge is very seldom achieved. 
Some approaches seek to develop reduced models of CRNs often based on eliminating the fast-evolving species~\cite{Segel1989, Lee2009, Gunawardena2012, Gunawardena2014}.
Other approaches such as flux balance analysis impose a complicated mix of constraints (physical and experimental) and objective functions (enforcing biologically desired results) to determine the currents through the CRN and avoid using kinetic information about the system \cite{Palsson2006,Palsson2011,Palsson2015} (see also Sec.~\ref{sec:discussion}).  
In both cases ensuring the thermodynamic consistency of the schemes has been a major topic of concern in recent years~\cite{Wachtel2018, Avanzini2020b,Beard2004,Kummel2006,Saldida2020,Akbari2021}. 

In this paper we present a novel approach:
a thermodynamically consistent circuit theory of CRNs, inspired by electronic circuit theory.
In CRNs elementary reactions transform chemical species into each other, while in electrical circuits devices transfer charges between conductors. 
But electronic devices are complex objects and the charge transfers are not characterized at an elementary level but instead in terms of current-voltage (I-V) curves which are often determined experimentally or may also be computed using a more detailed description of the inner workings of the device.
We do the same for CRNs. We group elementary reactions into \textit{chemical modules} that are then solely characterized by their current-concentration curves between \textit{terminal species}.
The current-concentration curve of a chemical module thus corresponds to the I-V curve of an electronic device, but differs from it in an important point.
While the electric currents only depend on the difference between the electrostatic potentials applied to the terminals of the devices, the chemical currents are functions of the concentrations and, consequently, depend on the absolute value of the chemical potentials of the terminal species.
Another difference between the two circuit theories is that conservation laws in CRNs are significantly more complicated than in electronic circuits where only charge conservation is involved.
Chemical circuit theory may become an important tool to study and design complex CRNs, in the same way that electronic circuit theory for electrical circuits has become the cornerstone of electrical engineering. To get there, experimental methodologies to determine current-concentration curves should be developed. This should be within reach thanks to recent developments in microfluidics and systems chemistry \cite{Hermans2017}.  

In order to present our theory, we adopt a two-fold strategy.
In the main text, we adopt an informal style and present the theory by examples, as we simplify the description of the CRN depicted in Fig.~\ref{fig:modularization}a into the one depicted in Fig.~\ref{fig:modularization}b,
first identifying the chemical modules (Sec.~\ref{sec:module_def}) and then characterizing them in terms of their current-concentration characteristic (Sec.~\ref{sec:currents}).
The formal theory is instead presented in App.~\ref{app:def_chem_module}.
The more mathematically inclined readers may want to start from there before turning to the main text for illustrations.

In Fig.~\ref{fig:modularization}a, the outer black box defines the boundary of the entire open CRN and the species with arrows crossing that boundary have their concentration controlled by the environment. 
The colored boxes inside the CRN denote the chemical modules and the corresponding colored arrows denote the (elementary  and reversible) reactions inside those modules. 
Note that the colored (internal) species change solely due to reactions in the module of the same color, while the black (terminal) species are involved in reactions of different modules. 
In Fig.~\ref{fig:modularization}b, the reactions within the modules are lumped into a minimal number of effective reactions called emergent cycles. As we will see, an emergent cycle defines a combination of elementary reactions that upon completion do not interconvert the {internal} species of a module, but exchange {terminal} species with other modules.
They have been originally introduced because they capture the entire dissipation of open CRNs at steady state~\cite{Polettini2014,Rao2016,Avanzini2021a}. 
The current along the emergent cycles of a module as a function of the concentrations of its terminal species defines the current-concentration curve of the module.
Three strategies (Sec.~\ref{sec:currents}) can be used to determine the current-concentration characteristic.
The first two (illustrated in App.~\ref{app:illustration_current_analytical} for some of the modules in Fig.~\ref{fig:modularization}a) are theoretical and require the detailed knowledge of the kinetic properties of the reactions inside the module. 
The third one (detailed in Sec.~\ref{sec:currents}) is experimental and requires the control of the concentrations of the terminal species as well as measuring their consumption/production rates. It is analogous to the way the I-V curve of an electronic device is determined.
Finally, based on the current-concentration characteristics of each module, a closed dynamics for the terminal species is obtained in Eq.~\eqref{eq:rate_eq_ct} providing a simplified description of the original open CRN.
Crucially, this coarse-grained dynamics is thermodynamically consistent  (Subs.~\ref{sub:thermo}) and  satisfies the chemical equivalent of Kirchhoff’s current and potential laws (Subs.~\ref{sub:kirchhoff}): 
the sums of currents involving each terminal species vanish at steady state and
the sum of the Gibbs free energy of reaction along any closed cycle is zero, respectively.
The limitations and extensions of our circuit theory are discussed in Sec.~\ref{sec:discussion} and illustrated in more detail in App.~\ref{app:limits} for the CRN in Fig.~\ref{fig:modularization}a.
To be valid beyond steady-state conditions, our theory requires a time-scale separation between the dynamics of terminal species and the internal dynamics of the modules in such a way that the latter is uniquely determined by the former, but multistability (Subs.~\ref{sub:instability}) can be treated anyway.
Modules may be merged into a super-module (Subs.~\ref{sub:open_as_module}) or split into submodules under certain conditions (Subs.~\ref{sub:decomposition}).
Finally, the effective reactions can be experimentally determined without knowing the internal stoichiometry of the modules (Subs.~\ref{sub:effrct_via_experiment}).

\section{Chemical Modules\label{sec:module_def}}
\begin{figure*}[t]
  \centering
    \includegraphics[width=0.99\columnwidth]{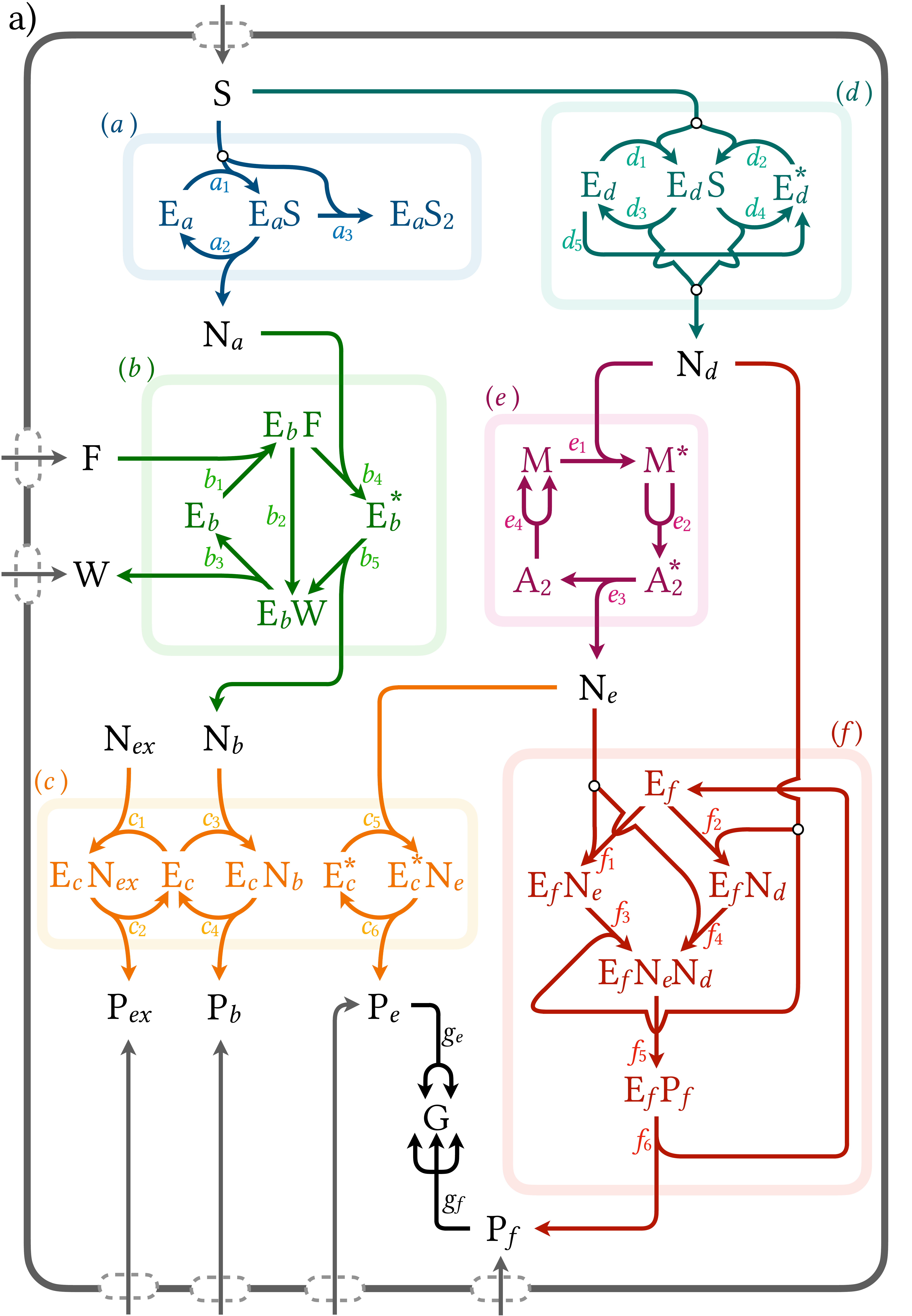}
  \hspace{2em}
    \includegraphics[width=0.99\columnwidth]{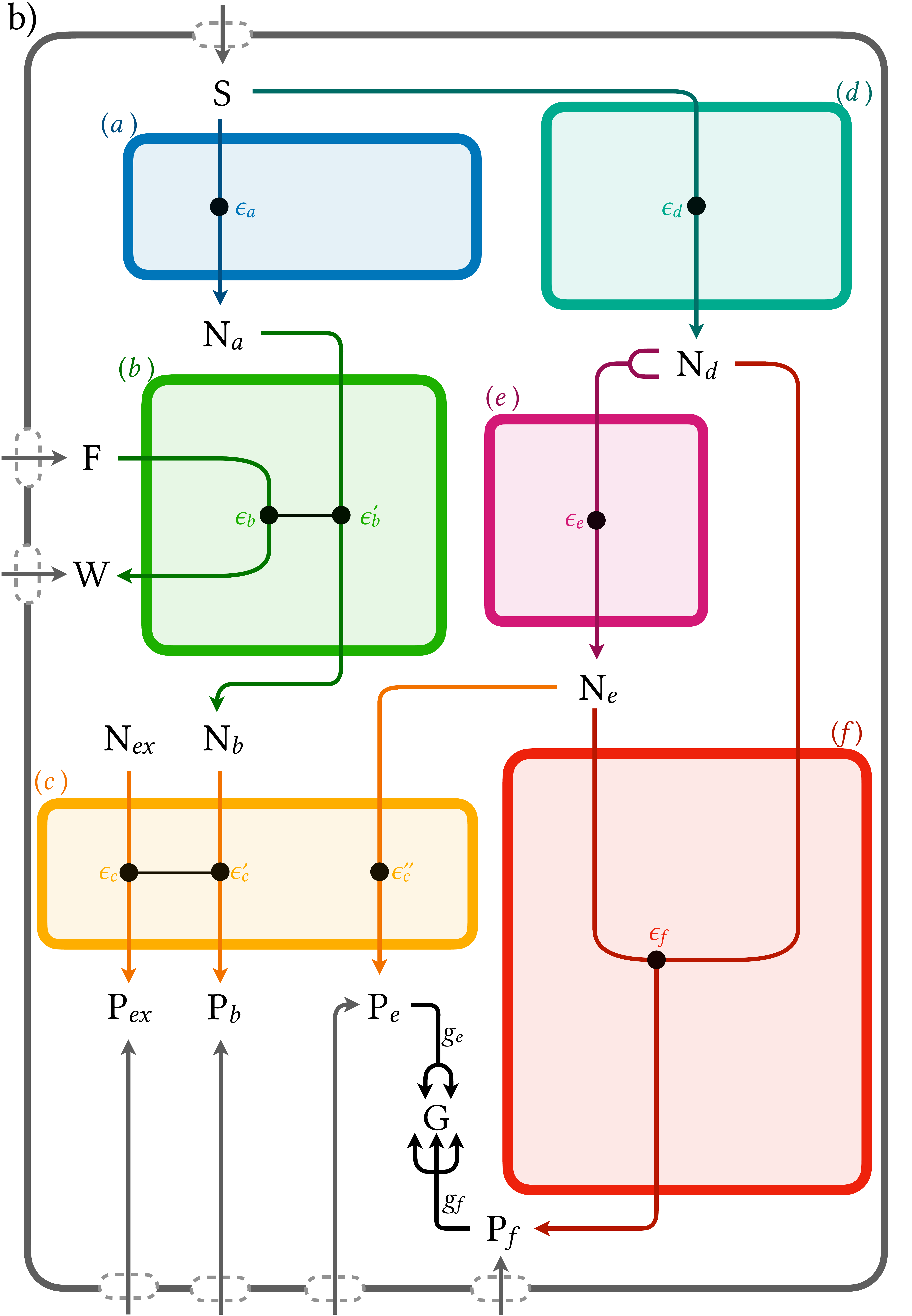}
\caption{
Elementary (a) and circuit description (b) of a complex CRN enclosed in gray boundaries using hypergraph notation.
Gray arrows crossing the network boundaries denote the exchange processes with the environment. 
Colored boxes are chemical modules.
Their internal species and (elementary or effective) reactions are represented by chemical symbols and arrows of the same color, respectively.
Their terminal species are represented by black chemical symbols.
All the chemical reactions are assumed to be reversible even though only the forward reactions are represented.
Effective reactions are coupled when connected by black lines.
}
\label{fig:modularization}
\end{figure*}

\begin{figure*}[p]
  \centering
  \includegraphics[width=1.9\columnwidth]{figure2.pdf}
\caption{
Chemical reactions, stoichiometric matrix, and cycles of the modules in Fig.~\ref{fig:modularization}a.
The black horizontal line splits the stoichiometric matrix of each module $\matSM$ into the substoichiometric matrix for the internal species $\matSQM$ and for the terminal species $\matSPM$. 
}
\label{fig:modularization_stoichiometry}
\end{figure*}

To explain how to reduce the description of a complex open CRN in terms of chemical modules, we will use the CRN depicted in Fig.~\ref{fig:modularization}a and reduce it to Fig.~\ref{fig:modularization}b. 
The formal description of this procedure is given in App.~\ref{app:def_chem_module}.
In particular, App.~\ref{app:def_chem_module_el} gives a formal definition of modules, while in Apps.~\ref{app:def_chem_module_eff} and~\ref{app:def_chem_module_eff_oscillations} their reduced description is derived.

In Fig.~\ref{fig:modularization}a, the arrows denote both the chemical reactions of the network as well as the exchange processes with the environment.
The latter are represented by (gray) arrows entering the CRN from the outside and involve the \textit{exchanged} species ($\ch{S}$, $\ch{F}$, $\ch{W}$, $\ch{P_{$ex$}}$, $\ch{P_{$\lmb$}}$, $\ch{P_{$\lme$}}$, and $\ch{P_{$\lmf$}}$).
The direction of the arrows is arbitrary (set by convention) as all reactions are assumed to be reversible. 
The boxes inside the CRN in Fig.~\ref{fig:modularization}a are the modules.
Each module is a subnetwork composed of a unique set of \textit{internal} species (drawn inside the module) reacting among themselves and potentially also with other species, named \textit{terminal} species (drawn outside the module).
For instance, the (blue) module~$\lma$ in Fig.~\ref{fig:modularization}a interconverts the internal species $\ch{E_{$\lma$}}\,,\text{ }\ch{E_{$\lma$}S}\,,\text{ }\ch{E_{$\lma$}S_2}$
and the terminal species $\ch{S}\text{ }\text{ and }\text{ }\ch{N_{$\lma$}}$
via the chemical reactions
\begin{equation}
\begin{split}
\ch{S + E_{$\lma$} &<=>[ $+\lma_{1}$ ][ $-\lma_{1}$ ] E_{$\lma$}S }\\
\ch{E_{$\lma$}S &<=>[ $+\lma_{2}$ ][ $-\lma_{2}$ ] E_{$\lma$} + N_{$\lma$} }\\
\ch{E_{$\lma$}S + S &<=>[ $+\lma_{3}$ ][ $-\lma_{3}$ ] E_{$\lma$}S_2 }
\end{split}
\label{eq:module1_rct}
\end{equation}
represented by the (blue) arrows labeled $\lma_1$, $\lma_2$, and $\lma_3$ (also specified in Fig.~\ref{fig:modularization_stoichiometry}).

Chemical modules are the chemical analog of the electronic components (for example diodes, transistors, or microchips) of an electric circuit, and the terminal species are the analog of the electrical contacts or pins of each component. 
Arrows in Fig.~\ref{fig:modularization}a should however not be compared to cables or connections between components in an electronic circuit diagram. Instead, the analog of electrical connections between contacts of different electronic components is the chemical species shared between chemical modules, i.e., the terminal species.
But while electronic components are spatially separated, chemical modules do not have to be.
Assuming homogeneous solutions for simplicity, the 
definition of the modules as well as their representation is based on the network of reactions and does not require any spatial organization.
Situations involving spatial organization will be discussed in Sec.~\ref{sec:discussion}.

In the circuit description depicted in Fig.~\ref{fig:modularization}b, each module ends up being coarse grained into effective reactions (denoted by the arrows through the boxes) between its {terminal} species.
The coarse graining reduces the (blue) module~$\lma$ to the single effective reaction
\begin{equation}
\ch{S <=>[ $\emer_\lma$ ][  ] N_{$\lma$}}\,.
\label{eq:module1_eff_rct}
\end{equation}
The coarse-graining procedure is based on the stoichiometry of the module and starts from its stoichiometric matrix~\cite{Palsson2011}
\begin{equation}
{\smatSM{\lma}}=
 \kbordermatrix{
   						 &\color{g}\lma_1 &\color{g}\lma_2&\color{g}\lma_3\cr
    \color{g}\ch{E_{$\lma$}}    		&-1	    	& 1	    	& 0 \cr
    \color{g}\ch{E_{$\lma$}S}  	    	& 1  		&-1	    	&-1\cr
    \color{g}\ch{E_{$\lma$}S_2} 	 	& 0  		& 0	    	& 1\cr \cline{2-4}
    \color{g}\ch{S}  			    	&-1 		& 0	    	&-1\cr
    \color{g}\ch{N_{$\lma$}} 	    	& 0  		& 1	    	& 0\cr
  }
  \begin{blockarray}{cc}
  &\\
  & \\
  \begin{block}{c\}\BAmultirow{1cm}}
  & \relax \\
  & \: $\mathbb{S}_a^Q$\\
  &\relax\\
  \end{block}
  \begin{block}{c\}\BAmultirow{1cm}}
  & \: $\mathbb{S}_a^P$\\
  & \relax\\
  \end{block}
  & \\
  \end{blockarray},
 \label{eq:module1_matS}
\end{equation}
also specified in Fig.~\ref{fig:modularization_stoichiometry}, whose entries have a clear physical meaning: they encode the net variation of the number of molecules of each species (identified by the matrix row) undergoing each reaction (identified by the matrix column)~\cite{Palsson2011}.
This matrix is split into the substoichiometric matrices for the internal species ${\smatSQM{\lma}}$ and for the terminal species ${\smatSPM{\lma}}$.
The effective reactions correspond to the \textit{emergent cycles} of ${\smatSM{\lma}}$, i.e., a set of linearly independent right-null vectors of ${\smatSQM{\lma}}$ that are not right-null vectors of ${\smatSPM{\lma}}$. As we will see, this set may not be unique. 
However, for the stoichiometric matrix~\eqref{eq:module1_matS}, one only finds the single emergent cycle:
\begin{equation}
{\cycle_{\emer_\lma}}=
 \kbordermatrix{
   						 &\cr
    \color{g}\lma_1    	&1		\cr
    \color{g}\lma_2 	&1  		\cr
    \color{g}\lma_3  	&0  		\cr
  }\,,
\label{eq:module1_emcycle}
\end{equation}
also reported in Fig.~\ref{fig:modularization_stoichiometry}.
The sequence of reactions encoded in the emergent cycle interconverts (upon completion) only the terminal species while leaving the internal species unaltered.
By multiplying the substoichiometry matrix for the terminal species ${\smatSPM{\lma}}$ in~\eqref{eq:module1_matS} and the emergent cycle ${\cycle_{\emer_\lma}}$ in~\eqref{eq:module1_emcycle}, one obtains the variation of the number of molecules of terminal species along the emergent cycle, i.e., the stoichiometry of the corresponding effective reaction~\eqref{eq:module1_eff_rct}:
\begin{equation}
{\smatSPM{\lma}} {\cycle_{\emer_\lma}}=
 \kbordermatrix{
   						 &\color{g}\cr
    \color{g}\ch{S}  			&-1 \cr
    \color{g}\ch{N_{$\lma$}} 		& 1 \cr
  }\,.
\end{equation}
The general and formal discussion of the coarse-grained procedure based on the use of the emergent cycles is given in App.~\ref{app:def_chem_module_eff} and App.~\ref{app:def_chem_module_eff_oscillations}.

We now turn to the (green) module~$\lmb$, whose internal species \ch{E_{$\lmb$}}, \ch{E_{$\lmb$}F}, \ch{E_{$\lmb$}W}, and \ch{E_{$\lmb$}^{*}} react via the chemical reactions $\lmb_1$, $\lmb_2$, $\lmb_3$, $\lmb_4$, and $\lmb_5$ (see Fig.~\ref{fig:modularization_stoichiometry})
with the terminal species \ch{N_{$\lma$}}, \ch{N_{$\lmb$}}, \ch{F}, and \ch{W}.
From the corresponding stoichiometric matrix $\smatSM{\lmb}$ (in Fig.~\ref{fig:modularization_stoichiometry}),
we identify two emergent cycles ${\cycle_{\emer_\lmb}}$ and ${\cycle_{\emer_\lmb'}}$ (in Fig.~\ref{fig:modularization_stoichiometry})
which correspond to the following effective reactions between the terminal species
\begin{subequations}
\begin{align}
\ch{F &<=>[ $\emer_\lmb$ ][ ] W }\label{eq:module2_eff_rct1}\,,\\
\ch{N_{$\lma$} &<=>[ $\emer_\lmb'$ ][ ] N_{$\lmb$} }\,,\label{eq:module2_eff_rct2}
\end{align}
\end{subequations}
respectively.
Note that 
\begin{equation}
{\cycle_{\emer_\lmb''}}=
 \kbordermatrix{
   			&\cr
    \color{g}\lmb_1    	& 1		\cr
    \color{g}\lmb_2 	& 0  		\cr
    \color{g}\lmb_3  	& 1  		\cr
    \color{g}\lmb_4  	& 1  		\cr
    \color{g}\lmb_5  	& 1  		\cr
  }
\end{equation}
is also a right-null vector of $\smatSQM{\lmb}$, which corresponds to the effective reaction
\begin{equation}
\ch{F + N_{$\lma$} <=>[ $\emer_\lmb''$ ][ ] N_{$\lmb$} + W }\,.
\end{equation}
But it is linearly dependent on the other two and is thus excluded from the circuit description. Any other pair of these three emergent cycles could also have been chosen.

We consider now the (aqua green) module~$\lmd$ whose internal species \ch{E_{$\lmd$}}, \ch{E_{$\lmd$}^{*}}, and \ch{E_{$\lmd$}S}  are involved in the chemical reactions $\lmd_1$, $\lmd_2$, $\lmd_3$, $\lmd_4$, and $\lmd_5$ (see Fig.~\ref{fig:modularization_stoichiometry})
with the terminal species \ch{S} and \ch{N_{$\lmd$}}.
Its stoichiometric matrix is specified in Fig.~\ref{fig:modularization_stoichiometry}
and, unlike the previous modules, the substoichiometry matrix for the internal species~${\smatSQM{\lmd}}$ admits the right-null vectors ${\cycle_{\inter_{\lmd}}}$ and ${\cycle_{\inter_{\lmd'}}}$, called \textit{internal cycles}, that are also right-null vectors of the substoichiometry matrix for the terminal species~${\smatSPM{\lmd}}$.
These internal cycles are sequences of reactions that upon completion leave all the species (both internal and terminal)  unaltered. 
Thus, they do not correspond to any effective reaction between terminal species.
However, the substoichiometry matrix for the internal species admits also the emergent cycle ${\cycle_{\emer_{\lmd}}}$
which corresponds to the following effective reaction 
\begin{equation}
\ch{S<=>[ $\emer_\lmd$ ][ $$ ] N_{$\lmd$} }\,.
\label{eq:module4_eff_rct}
\end{equation}

By following the same procedure for the remaining modules, one obtains the following effective reactions
\begin{equation}
\begin{split}
\ch{N_{$ex$} &<=>[ $\emer_\lmc$ ][ $$ ] P_{$ex$} }\\
\ch{N_{$\lmb$} &<=>[ $\emer_\lmc'$ ][ $$ ] P_{$\lmb$} }\\
\ch{N_{$\lme$} &<=>[ $\emer_\lmc''$ ][ $$ ] P_{$\lme$} }
\end{split}
\label{eq:module3_eff_rct}
\end{equation}
for the (orange) module~$\lmc$;
\begin{equation}
\ch{2 N_{$\lmd$} <=>[ $\emer_\lme$ ][ $$ ] N_{$\lme$}}\,,
\label{eq:module5_eff_rct}
\end{equation}
for the (purple) module~$\lme$; and
\begin{equation}
\ch{ N_{$\lmd$} + N_{$\lme$} <=>[ $\emer_\lmf$ ][ $$ ] P_{$\lmf$}}\,.
\label{eq:module6_eff_rct}
\end{equation}
for the (red) module~$\lmf$.

Note that a chemical module $\module$ with $\nterminalm$ terminal species can have a maximum of $\nterminalm-1$ (linearly independent) emergent cycles and, therefore, effective reactions. 
This is analogous to the fact that an electronic component with $\nterminalm$ contacts can have at most $\nterminalm-1$ independent electrical currents at steady state~\cite{Freitas2021a}. 
This follows directly from the existence of at least one conservation law, namely, mass conservation law in CRNs or electric charge conservation law in electronic circuits.
The existence of additional conservation laws (involving the terminal species in CRNs or the contacts in electronic circuits) reduces the number of emergent cycles. 
In the case of electronic circuits, the only kind of conservation law is the charge, and the only way to have additional conservation laws beyond that of the total charge is for a component to consist of smaller subcomponents that do not interchange any charge (although they might still influence each other). 
This is not the case for CRNs, where conservation laws (involving the terminal species) can be more complicated~\cite{Polettini2014,Rao2016}. 
They  identify parts of (or entire) molecules, named moieties, that are not modified by the chemical reactions. 
Mathematically, they correspond to left-null vectors of the full stoichiometric matrix $\matSM$ of a module whose restrictions to the internal species are not left-null vectors of the stoichiometric matrix $\matSQM$ for the internal species. 
It can be seen that if $\nbrokenm$ and $\nemerm$ are the number of independent conservation laws and emergent cycles, respectively, then $\nbrokenm+\nemerm=\nterminalm$~\cite{Rao2016}.
This, together with the existence of at least one conservation law (mass conservation law in CRNs or charge conservation in electronic circuits), 
explains why the number of emergent cycles is at most $\nterminalm-1$.
Note that these conservation laws are said to be broken because they define quantities that are only conserved in the closed system (CRNs~\cite{Rao2016} or electronic circuits~\cite{Freitas2021a}).


\section{Current-Concentration Characteristic\label{sec:currents}}

In electronic circuits, the steady-state behavior of electronic components is given by their current-voltage characteristics, or ``I-V curves'', which specify how the value of all independent currents of an electronic component depends on the voltages applied to its contacts.
We now apply the same strategy to chemical modules.
The current-concentration characteristic of a chemical module specifies how the (effective) reaction currents depend on the concentrations of the terminal species only,
by assuming that the internal species have already relaxed to steady state (see App.~\ref{app:def_chem_module_eff}). 

When the kinetic constants of the internal reactions of a module are known, the current-concentration characteristic can be derived analytically if the internal reactions of a module are pseudo-first-order reactions, or otherwise numerically. 
The procedures to do so are respectively described in App.~\ref{sub:effcurr_analytical} and App.~\ref{sub:effcurr_numerical} and applied to some of the modules in Fig.~\ref{fig:modularization}a.
However, in practice, a complete characterization of the kinetics of the internal reactions is seldom achieved. 
The real power of the circuit theory is that the current-concentration characteristic can be determined experimentally, as discussed next.

\begin{figure}[t]
  \centering
  \includegraphics[width=0.99\columnwidth]{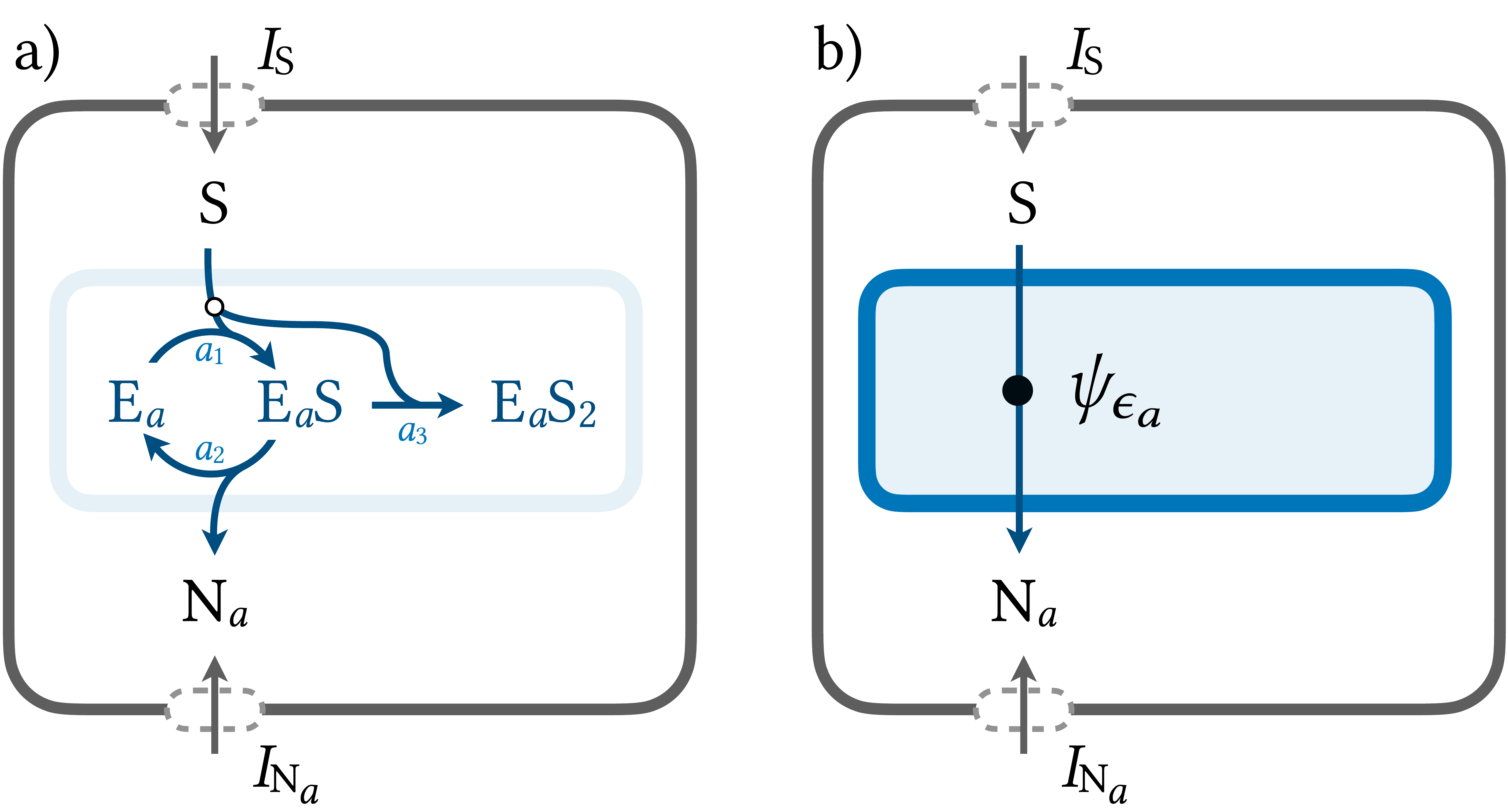}
\caption{Elementary (a) and circuit description (b) of the (blue) module~$a$ in Fig.~\ref{fig:modularization} in a reactor, similar to the membrane reactor used in Ref.~\cite{Sorrenti2017}, where the concentrations of \ch{S} and \ch{N_{$\lma$}} are controlled by exchange processes whose currents are specified by $ \ssexcurrel_{\ch{S}}$ and $\ssexcurrel_{\ch{N_{$\lma$}}}$.}
\label{fig:IVcharacteristicA}
\end{figure}

One possible way may be to resort to membrane reactors~\cite{Sorrenti2017}.
We describe the procedure using the (blue) module~$\lma$ and the (green) module~$\lmb$ in Fig.~\ref{fig:modularization}a. The formal theory is detailed in App.~\ref{app:definition_current_experimental}.
The setup to characterize the (blue) module~$\lma$ is illustrated in Fig.~\ref{fig:IVcharacteristicA}. 
The concentrations of \ch{S} and \ch{N_{$\lma$}} are held constant thanks to the exchange processes whose currents are $\ssexcurrel_{\ch{S}}$ and $\ssexcurrel_{\ch{N_{$\lma$}}}$ satisfying:
\begin{subequations}
\begin{align}
\dt[\ch{S}]&=-\smodcyclecurrel{\emer_\lma}+\ssexcurrel_{\ch{S}}=0\,,\\
\dt[\ch{N_{$\lma$}}]&=\smodcyclecurrel{\emer_\lma}+\ssexcurrel_{\ch{N_{$\lma$}}}=0\,.
\end{align}
\end{subequations}
Thus, the effective reaction current $\smodcyclecurrel{\emer_\lma}$ can be determined by measuring the exchange current $\ssexcurrel_{\ch{S}}$ (or equivalently $\ssexcurrel_{\ch{N_{$\lma$}}}$) for every value of the concentrations $[\ch{S}]$ and $[\ch{N_{$\lma$}}]$: 
\begin{equation}
\smodcyclecurrel{\emer_\lma} = - \ssexcurrel_{\ch{S}}  = \ssexcurrel_{\ch{N_{$\lma$}}}\,.
\end{equation}

\begin{figure}[t]
  \centering
  \includegraphics[width=0.99\columnwidth]{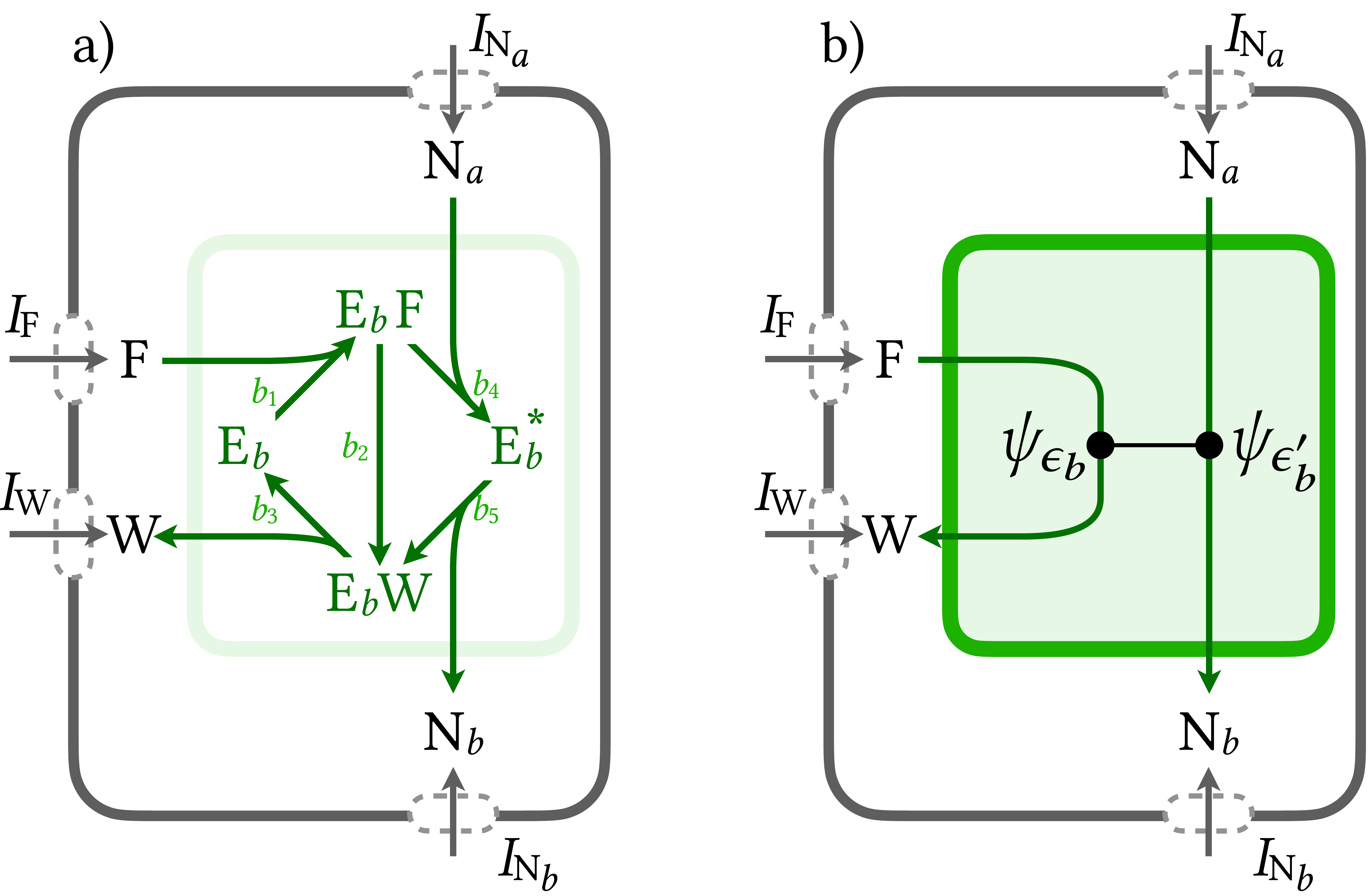}
\caption{Elementary (a) and circuit description (b) of the (green) module~$b$ in  Fig.~\ref{fig:modularization} in a reactor, similar to the membrane reactor used in Ref.~\cite{Sorrenti2017}, where the concentrations of \ch{N_{$\lma$}}, \ch{N_{$\lmb$}}, \ch{F} and \ch{W} are controlled by exchange processes whose currents are specified by  $ \ssexcurrel_{\ch{N_{$\lma$}}}$, $\ssexcurrel_{\ch{N_{$\lmb$}}}$, $\ssexcurrel_{\ch{F}}$, and $\ssexcurrel_{\ch{W}}$.}
\label{fig:IVcharacteristicB}
\end{figure}

The setup for the (green) module~$\lmb$ in Fig.~\ref{fig:modularization}a is illustrated in Fig.~\ref{fig:IVcharacteristicB}. 
The module has now two effective reactions,~\eqref{eq:module2_eff_rct1} and~\eqref{eq:module2_eff_rct2}, but the general strategy remains the same. The concentrations of the terminal species $(\ch{N_{$\lma$}},\ch{N_{$\lmb$}}, \ch{F},  \ch{W})$ are held constant thanks to the exchange processes whose currents are $\ssexcurrel_{\ch{N_{$\lma$}}}$,  $\ssexcurrel_{\ch{N_{$\lmb$}}}$, $\ssexcurrel_{\ch{F}}$, and $\ssexcurrel_{\ch{W}}$ satisfying:
\begin{subequations}
\begin{align}
\dt[\ch{N_{$\lma$}}]&=-\smodcyclecurrel{\emer_\lmb'}+\ssexcurrel_{\ch{N_{$\lma$}}}=0\,,\\
\dt[\ch{N_{$\lmb$}}]&=\smodcyclecurrel{\emer_\lmb'}+\ssexcurrel_{\ch{N_{$\lmb$}}}=0\,,\\
\dt[\ch{F}]&=-\smodcyclecurrel{\emer_\lmb}+\ssexcurrel_{\ch{F}}=0\,,\\
\dt[\ch{W}]&=\smodcyclecurrel{\emer_\lmb}+\ssexcurrel_{\ch{W}}=0\,.
\end{align}
\end{subequations}
and thus the effective reaction currents~$\smodcyclecurrel{\emer_\lmb}$ and$\smodcyclecurrel{\emer_\lmb'}$ are given by
\begin{subequations}
\begin{align}
\smodcyclecurrel{\emer_\lmb'}&=\ssexcurrel_{\ch{N_{$\lma$}}}=-\ssexcurrel_{\ch{N_{$\lmb$}}}\,,\\
\smodcyclecurrel{\emer_\lmb}&=\ssexcurrel_{\ch{F}}=-\ssexcurrel_{\ch{W}}\,.
\end{align}
\end{subequations}
This operation can be repeated for every module.
In App.~\ref{app:definition_current_experimental}, we formally derive 
the general expression (Eq.~\eqref{eq:cyclecurr2}) of the effective reaction currents in terms of the exchange currents.

\section{Circuit Description\label{sec:circuit_description}}

Having determined the effective reactions and the current-concentration characteristic of each module, we can finally formulate the circuit description of the CRN in Fig.~\ref{fig:modularization}b. 
The general formulation is presented in App.~\ref{app:kirchhoff}.

To do so, modules are connected by sharing their terminal species. 
For instance, the terminal species \ch{S} is involved in the two effective reactions $\emer_\lma$ and $\emer_\lmd$ (given in~\eqref{eq:module1_eff_rct} and~\eqref{eq:module4_eff_rct}, respectively) as a reagent and also exchanged with the environment.
Its concentration thus evolves according to
\begin{equation}
\dt[\ch{S}] = -\smodcyclecurrel{\emer_\lma} -\smodcyclecurrel{\emer_\lmd} + \excurrel_{\ch{S}}\,,
\end{equation}
where $\excurrel_{\ch{S}}$ is the exchange current of \ch{S} with the environment.
Analogously, the terminal species \ch{N_{$\lmd$}} is the product of the effective reaction $\emer_\lmd$ (given in~\eqref{eq:module4_eff_rct}) and the reagent of the effective reactions $\emer_\lme$ and $\emer_\lmf$ (given in~\eqref{eq:module5_eff_rct} and~\eqref{eq:module6_eff_rct}, respectively).
Its concentration thus follows
\begin{equation}
\dt[\ch{N_{$\lmd$}}] = \smodcyclecurrel{\emer_\lmd} -2\smodcyclecurrel{\emer_\lme} - \smodcyclecurrel{\emer_\lmf}\,,
\end{equation}
which accounts for the fact that 2 molecules of \ch{N_{$\lmd$}} are consumed every time reaction $\emer_\lme$ occurs, namely, the stoichiometry of the effective reaction $\emer_\lme$.

By repeating the same reasoning, the rate equation for the terminal species can be written as 
\begin{equation}
\dt \terconc = \matSeff {\cyclecurr} + \excurr\,,
\label{eq:rate_eq_ct}  
\end{equation}
where 
{\small\begin{equation}
\matSeff=
 \kbordermatrix{
&\color{g}\emer_{\lma}&\color{g}\emer_{\lmb}&\color{g}\emer_{\lmb}'&\color{g}\emer_{\lmc}&\color{g}\emer_{\lmc}'&\color{g}\emer_{\lmc}''&\color{g}\emer_{\lmd}&\color{g}\emer_{\lme}&\color{g}\emer_{\lmf}&\color{g}\lmge&\color{g}\lmgf\cr
    \color{g}\ch{N_{$\lma$}}    	&1		&0		&-1		&0		&0		&0		&0		&0		&0		&0		&0\cr
    \color{g}\ch{N_{$\lmb$}}    	&0		&0		&1		&0		&-1		&0		&0		&0		&0		&0		&0\cr
    \color{g}\ch{N_{$ex$}}    	&0		&0		&0		&-1		&0		&0		&0		&0		&0		&0		&0\cr
    \color{g}\ch{N_{$\lmd$}}    	&0		&0		&0		&0		&0		&0		&1		&-2		&-1		&0		&0\cr
    \color{g}\ch{N_{$\lme$}}    	&0		&0		&0		&0		&0		&-1		&0		&1		&-1		&0		&0\cr
    \color{g}\ch{G}   		 	&0		&0		&0		&0		&0		&0		&0		&0		&0		&2		&3\cr
    \cline{2-12}
    \color{g}\ch{F}   		 	&0		&-1		&0		&0		&0		&0		&0		&0		&0		&0		&0\cr
    \color{g}\ch{W}  		  	&0		&1		&0		&0		&0		&0		&0		&0		&0		&0		&0\cr
    \color{g}\ch{S}   		 	&-1		&0		&0		&0		&0		&0		&-1		&0		&0		&0		&0\cr
    \color{g}\ch{P_{$\lmb$}}    	&0		&0		&0		&0		&1		&0		&0		&0		&0		&0		&0\cr
    \color{g}\ch{P_{$ex$}}    	&0		&0		&0		&1		&0		&0		&0		&0		&0		&0		&0\cr
    \color{g}\ch{P_{$\lme$}}    	&0		&0		&0		&0		&0		&1		&0		&0		&0		&-1		&0\cr
    \color{g}\ch{P_{$\lmf$}}    	&0		&0		&0		&0		&0		&0		&0		&0		&1		&0		&-1\cr
 }\,
 \label{eq:stoichiometric_circuit}
\end{equation}}
is the stoichiometric matrix of the effective reactions (where the black horizontal line splits the set of species into internal and exchanged species), and
{\small\begin{align}
\terconc = 
 \begin{pmatrix}
    [\ch{N_{$\lma$}}] 	\\
    [\ch{N_{$\lmb$}}]  \\
    [\ch{N_{$ex$}}]  \\
    [\ch{N_{$\lmd$}}]	\\
    [\ch{N_{$\lme$}}]\\
    [\ch{G}]	\\
    [\ch{F}] \\
    [\ch{W}]\\
    [\ch{S}]\\
    [\ch{P_{$\lmb$}}] \\
    [\ch{P_{$ex$}}] \\
    [\ch{P_{$\lme$}}]\\
    [\ch{P_{$\lmf$}}]  
  \end{pmatrix}\,,
 \text{ }\text{ }\text{ }\text{ }\text{ }\text{ }\text{ }
\cyclecurr = 
 \begin{pmatrix}
 \modcyclecurrel_{\emer_{\lma}}\\
 \modcyclecurrel_{\emer_{\lmb}}\\
 \modcyclecurrel_{\emer_{\lmb}'}\\
 \modcyclecurrel_{\emer_{\lmc}}\\
 \modcyclecurrel_{\emer_{\lmc}'}\\
 \modcyclecurrel_{\emer_{\lmc}''}\\
 \modcyclecurrel_{\emer_{\lmd}}\\
 \modcyclecurrel_{\emer_{\lme}}\\
 \modcyclecurrel_{\emer_{\lmf}}\\
 \modcyclecurrel_{{\lmge}}\\
 \modcyclecurrel_{{\lmgf}}
 \end{pmatrix}\,,
 \text{ }\text{ }\text{ }\text{ and }\text{ }\text{ }\text{ }
 \excurr = 
 \begin{pmatrix}
    0 \\
    0 \\
    0 \\
    0 \\
    0 \\
    0 \\
    \excurrel_{\ch{F}} \\
    \excurrel_{\ch{W}} \\
    \excurrel_{\ch{S}} \\
    \excurrel_{\ch{P_{$\lmb$}}} \\
    \excurrel_{\ch{P_{$ex$}}} \\
    \excurrel_{\ch{P_{$\lme$}}}\\
    \excurrel_{\ch{P_{$\lmf$}}} 
  \end{pmatrix}\,
  \label{eq:curr_CD}
\end{align}}
are the concentration vector of all the terminal species, the reaction current vector, and the exchange current vector, respectively.
Note that in $\cyclecurr$, the currents $\modcyclecurrel_{{\lmge}}$ and $\modcyclecurrel_{{\lmgf}}$ still correspond to elementary reactions 
\begin{subequations}
\begin{align}
\ch{P_{$\lme$} & <=>[ $+\lmge$ ][ $-\lmge$ ] 2 G}\,\\
\ch{P_{$\lmf$} &<=>[ $+\lmgf$ ][ $-\lmgf$ ] 3 G}\,,
\end{align}
\end{subequations}
of the complex CRN in Fig.~\ref{fig:modularization}a.

\subsection{Thermodynamics\label{sub:thermo}}

We emphasize that our circuit theory is thermodynamically consistent, contrary to many other coarse-graining schemes. 
This means that the entropy production rate of the coarse-grained description is the same as the entropy production of the original full description of the CRN. In other words, the reduction scheme preserves the entropy production rate of the CRN.

The chemical potential of a chemical species $\chemspecies$ in a homogeneous solution~\cite{Alberty2003} is given by
\begin{equation}
\chempotential_\chemspecies  = \stchempotential_\chemspecies+ RT \ln[\chemspecies]
\label{eq:chempot}
\end{equation}
where $\stchempotential_\chemspecies$ is the standard chemical potential, $R$ is the gas constant, and $T$ is the temperature of the solution. 
The Gibbs free energy change in a homogeneous CRN caused by a reaction~$\elrct$ is given by
\begin{equation}
\freerct{\elrct}=\sum_{\chemspecies}\chempotential_\chemspecies \elS^\chemspecies_\elrct\,,
\label{eq:deltaG}
\end{equation}
where $\elS^\chemspecies_\elrct$ is the net stoichiometric coefficient of the $\chemspecies$ species in the $\elrct$ reaction.
 For example, the Gibbs free energy changes of the internal reaction of the  (blue) module~$\lma$ in Fig.~\ref{fig:modularization}a are given by
\begin{subequations}
    \begin{align}
         \freerct{\lma_{1}} & = \chempotential_{\ch{E_{$\lma$}S}} - \chempotential_{\ch{ E_{$\lma$}}} - \chempotential_{\ch{S}}\,, \\
         \freerct{\lma_{2}} & = \chempotential_{\ch{E_{$\lma$}}} + \chempotential_{\ch{ N_{$\lma$}}} - \chempotential_{\ch{E_{$\lma$}S}}\,, \\
         \freerct{\lma_{3}} & = \chempotential_{\ch{E_{$\lma$}S_2}} - \chempotential_{\ch{E_{$\lma$}S}} - \chempotential_{\ch{S}}\,.
    \end{align}
\end{subequations}

At the elementary level of description of CRNs, the local detailed balance property 
\begin{equation}
-\freerct{\elrct}=RT\ln\frac{\currel_{+\elrct}}{\currel_{-\elrct}}
\label{eq:ldb}
\end{equation}
must hold. It relates thermodynamics to the log ratio of the forward and backward reaction fluxes $\currel_{\pm\elrct}$ (given by Eq.~\eqref{eq_mass_action}) contributing to the reaction currents $\currel_{\elrct}=\currel_{+\elrct} - \currel_{-\elrct}$. 
Furthermore, the entropy production rate of elementary CRNs (also called the total dissipation) reads
\begin{equation}
T\epr = -\sum_{\elrct}\freerct{\elrct}\,\currel_{\elrct} \, ,
\label{eq:epr}
\end{equation}
and quantifies the entropy change per unit of time in CRNs as well as in the thermal and chemical reservoirs~\cite{Rao2016}.
Together with the local detailed balance property \eqref{eq:ldb}, the entropy production rate can be rewritten in a manifestly non-negative form
\begin{equation}
T\epr = RT \sum_{\elrct}\,(\currel_{+\elrct}-\currel_{-\elrct})
\ln\frac{\currel_{+\elrct}}{\currel_{-\elrct}} \geq 0 \,,
\label{eq:eprBis}
\end{equation}
thus mathematically ensuring the validity of the second law.
The summation over $\elrct$ in Eq.~\eqref{eq:epr} runs over all the reactions excluding the exchange processes with the environment (i.e., all arrows but the grey ones in Fig.~\ref{fig:modularization}a).
Vanishing entropy production defines thermodynamic equilibrium where all reaction currents vanish. Together Eqs.~\eqref{eq:ldb}-\eqref{eq:eprBis} ensure a thermodynamically consistent descriptions of elementary CRNs.

Our circuit theory allows for a thermodynamically consistent description of CRNs made of effective reactions because the entropy production rate~\eqref{eq:epr} can be expressed as 
\begin{equation}
T\epr = -\sum_{\emer}\freerct{\emer}\,\modcyclecurrel_{\emer}\,,
\label{eq:epr_cg}
\end{equation}
where the summation now only runs over the effective reactions $\emer$ of the modules (all arrows but the grey ones in Fig.~\ref{fig:modularization}b) and the free energy along the effective reaction $\emer$ is given by
\begin{equation}
\freerct{\emer}=\sum_{\chemspecies\in\setterminal}\chempotential_\chemspecies \elSeff^\chemspecies_\emer\,,
\end{equation}
where $\setterminal$ is the set of all the terminal species.
For instance, the Gibbs free energy change along the single effective reaction of the (blue) module~$\lma$ reads
\begin{equation}
    \freerct{\emer_\lma} = \chempotential_{\ch{ N_{$\lma$}}} - \chempotential_{\ch{S}}\,.
\end{equation}
The remarkable reduction from \eqref{eq:epr} to \eqref{eq:epr_cg} arises because the emergent cycles define a minimal set of effective reactions which preserve the exact evolution of the terminal species while carrying the full dissipation of modules, as long as there is a time-scale separation between the dynamics of the internal and terminal species. This has been proven in Ref.~\cite{Avanzini2020b}.
We emphasize, however, that the entropy production rate at the circuit level \eqref{eq:epr_cg} cannot be, in general, expressed in a form reminiscent of \eqref{eq:eprBis} as shown in Ref.~\cite{Wachtel2018}.  
We also note that, unlike our theory, most coarse-graining schemes underestimate the exact dissipation even in presence of a time-scale separation (see for instance Ref.~\cite{Esposito2012}) because the fast degrees of freedom which have been eliminated are still out-of-equilibrium and contribute to the dissipation.

\subsection{Kirchhoff’s Laws\label{sub:kirchhoff}}

We now show that the dynamics of our circuit theory~\eqref{eq:rate_eq_ct} satisfies the chemical equivalent of Kirchhoff’s laws in electrical circuits. Here, we present these laws for the CRN in Fig.~\ref{fig:modularization}b, while their general formulation and derivation are given in App.~\ref{app:kirchhoff}.
We emphasize that Kirchhoff’s laws for CRNs at the level of the elementary dynamics are not new (see for instance Ref.~\cite{Oster1973, Beard2002, Qian2005, DalCengio2022}). The novelty of our approach is that Kirchhoff’s laws are recovered at the coarse-grained/circuit level.

Kirchhoff’s current law states that, at steady state, the sum of the currents entering into a node of an electronic circuit is equal to the sum of the currents exiting it. 
The terminal species correspond to the nodes of an electrical circuit in our circuit theory.
Hence, Kirchhoff’s current law can be expressed for the CRN in Fig.~\ref{fig:modularization}b in terms of the steady-state conditions (denoted by the overline) 
\begin{subequations}{\small
\begin{align}
    \dt\steady{[\ch{N_{$\lma$}}]} 	&= \ssmodcyclecurrel_{\emer_{\lma}} - \ssmodcyclecurrel_{\emer_{\lmb}'} 								= 0 \\
    \dt\steady{[\ch{N_{$\lmb$}}]} 	&= \ssmodcyclecurrel_{\emer_{\lmb}'} - \ssmodcyclecurrel_{\emer_{\lmc}'}								= 0 \\
    \dt\steady{[\ch{N_{$ex$}}]} 	&= -\ssmodcyclecurrel_{\emer_{\lmc}}														= 0 \\
    \dt\steady{[\ch{N_{$\lmd$}}]} 	&= \ssmodcyclecurrel_{\emer_{\lmd}} -2 \ssmodcyclecurrel_{\emer_{\lme}} - \ssmodcyclecurrel_{\emer_{\lmf}}	= 0 \\
    \dt\steady{[\ch{N_{$\lme$}}]} 	&= -\ssmodcyclecurrel_{\emer_{\lmc}''} + \ssmodcyclecurrel_{\emer_{\lme}} - \ssmodcyclecurrel_{\emer_{\lmf}}	= 0 \\
    \dt\steady{[\ch{G}]}		&= 2 \ssmodcyclecurrel_{{\lmge}} + 3 \ssmodcyclecurrel_{{\lmgf}} 									= 0 \\
    \dt\steady{[\ch{F}]}  		&=  -\ssmodcyclecurrel_{\emer_{\lmb}} + \ssssexcurrel_{\ch{F}} 									= 0 \\
    \dt\steady{[\ch{W}]} 		&=  \ssmodcyclecurrel_{\emer_{\lmb}} + \ssssexcurrel_{\ch{W}}										= 0 \\
    \dt\steady{[\ch{S}]}			&= - \ssmodcyclecurrel_{\emer_{\lma}} - \ssmodcyclecurrel_{\emer_{\lmd}} + \ssssexcurrel_{\ch{S}} 			= 0 \\
    \dt\steady{[\ch{P_{$\lmb$}}]} 	&=  \ssmodcyclecurrel_{\emer_{\lmc}'} + \ssssexcurrel_{\ch{P_{$\lmb$}}} 								= 0 \\
    \dt\steady{[\ch{P_{$ex$}}]} 	&= \ssmodcyclecurrel_{\emer_{\lmc}} + \ssssexcurrel_{\ch{P_{$ex$}}} 								= 0 \\
    \dt\steady{[\ch{P_{$\lme$}}]} 	&= \ssmodcyclecurrel_{\emer_{\lmc}''} -  \ssmodcyclecurrel_{{\lmge}} + \ssssexcurrel_{\ch{P_{$\lme$}}} 		= 0 \\
    \dt\steady{[\ch{P_{$\lmf$}}]} 	&= \ssmodcyclecurrel_{\emer_{\lmf}} - \ssmodcyclecurrel_{{\lmgf}} + \ssssexcurrel_{\ch{P_{$\lmf$}}} 			= 0
\end{align}}
\end{subequations}
imposing that the sum of the currents (both effective and exchange) affecting the concentration of each terminal species vanishes. 
This is formally derived by imposing that the left-hand-sides of Eq.~\eqref{eq:rate_eq_ct} vanish and by using the stoichiometric matrix given in Eq.~\eqref{eq:stoichiometric_circuit} and the currents given in Eq.~\eqref{eq:curr_CD}.

On the other hand, Kirchhoff’s potential law states that the sum of potential differences along any closed loop is zero.
In our circuit description of CRNs, loops correspond to the internal cycles of $\matSeff$ (introduced in Sec.~\ref{sec:module_def} and detailed in App.~\ref{app:def_chem_module_eff}) and potential differences to the variations of the Gibbs free energy along the (effective or not) reactions (e.g., $\{\freerct{\emer_{\lma}},\freerct{\emer_{\lmb}}, \freerct{\emer_{\lmb'}},\dots\}$ for the effective reactions in Fig.~\ref{fig:modularization}b). 
Since, the stoichiometric matrix~\eqref{eq:stoichiometric_circuit} admits only one internal cycle,
\begin{equation}
\cyclect_{\inter}=
 \kbordermatrix{
   						 &\cr
    \color{g}{\emer_{\lma}}    	& 0		\cr
    \color{g}{\emer_{\lmb}} 		& 0  		\cr
    \color{g}{\emer_{\lmb}'}  	& 0  		\cr
    \color{g}{\emer_{\lmc}}	  	& 0  		\cr
    \color{g}{\emer_{\lmc}'}	  	& 0  		\cr
    \color{g}{\emer_{\lmc}''}	  	&-3  		\cr
    \color{g}{\emer_{\lmd}}	  	& 0  		\cr
    \color{g}{\emer_{\lme}}	  	&-1  		\cr
    \color{g}{\emer_{\lmf}}	  	& 2  		\cr
    \color{g}{{\lmge}}		  	&-3  		\cr
    \color{g} {{\lmgf}}		  	& 2  		\cr
  }\,,
\label{eq:internal_CD} 
\end{equation}
Kirchhoff’s potential law can be expressed as
\begin{equation}
2\freerct{\emer_{\lmf}}+2\freerct{{\lmgf}}-3\freerct{{\lmge}}-3\freerct{\emer_{\lmc''}}-\freerct{\emer_{\lme}} = 0\,,
\label{eq:Kirchhoff_potential_law}
\end{equation}
for the CRN in Fig.~\ref{fig:modularization}b,
which using Eq.~\eqref{eq:deltaG} is indeed true since
\begin{subequations}
\begin{align}
\freerct{\emer_{\lmf}} & = \chempotential_{\ch{P}_\lmf} - \chempotential_{\ch{N}_{\lme}} - \chempotential_{\ch{N}_{\lmd}}\,, \\
\freerct{{\lmgf}} & = 3\chempotential_{\ch{G}} - \chempotential_{\ch{P}_{\lmf}}\,,\\
\freerct{{\lmge}} & = 2 \chempotential_{\ch{G}} - \chempotential_{\ch{P}_{\lme}}\,,\\
\freerct{\emer_{\lmc''}} & = \chempotential_{\ch{P}_{\lme}} - \chempotential_{\ch{N}_\lme}\,,\\
\freerct{\emer_{\lme}} & = \chempotential_{\ch{N}_\lme} - 2 \chempotential_{\ch{N}_\lmd}\,.
\end{align}
\end{subequations}
This is formally derived by imposing that 
the sum of the variations of the Gibbs free energy along the (effective or not) reactions (e.g., $\{\freerct{\emer_{\lma}},\freerct{\emer_{\lmb}}, \freerct{\emer_{\lmb'}},\dots\}$ multiplied by the corresponding entry of the internal cycles in Eq.~\eqref{eq:internal_CD} vanishes.

\section{Discussion and Perspectives \label{sec:discussion}}

We start by discussing how apparent limitations of our circuit theory may be overcome.

The fact that the current-concentration characteristic of a chemical module is evaluated assuming that the module is in a steady state (based on the time scale separation assumption mentioned before and formally discussed in App.~\ref{app:def_chem_module_eff}) may give the impression that oscillations in the concentrations of the species internal to a module compromise the theory. However, we prove in App.~\ref{app:def_chem_module_eff_oscillations} that this is not the case and that such oscillations can be treated as long as their period is much shorter than the time scale of the terminal species dynamics.

In certain situations, the current-concentration characteristic may be such that the effective reaction currents are not uniquely defined in terms of the concentrations of the terminal species. This will happen for modules with nonlinear chemical reactions displaying multistability.
In such cases, hysteresis effects may arise creating a dependence on the past history of the network, but the network theory is still applicable. An explicit example of such a situation is worked out in App.~\ref{sub:instability}.  

We presented the circuit theory starting from elementary reactions that we grouped into modules.
But naturally, modules can be further grouped into higher-level modules. We examine this in App.~\ref{sub:open_as_module} by showing that the entire CRN depicted in Fig.~\ref{fig:modularization}a/b can be treated as a module and its exchanged species become terminal species.
This also raises the question of what are the conditions under which a module can be decomposed into smaller modules. The answer is quite simple: as long as the effective reactions belonging to a smaller module are independent of those of another module, i.e., when their emergent cycles do not share internal species. Such a decomposition is discussed in App.~\ref{sub:decomposition} for the (orange) module~$\lmc$ in Fig.~\ref{fig:modularization}a.

When discussing the experimental characterization of the current-concentration characteristic, we implicitly assumed that the stoichiometry of the internal reactions of the module is known. However, even when this is not the case, recovering that stoichiometry is not too complicated experimentally. We illustrate how such a procedure might be implemented in App.~\ref{sub:effrct_via_experiment} for some of the modules in Fig.~\ref{fig:modularization}a.
  
Our circuit theory was presented here for ideal homogeneous solutions, but these conditions can easily be relaxed. 
Non-ideal solutions can be treated within mean-field theories~\cite{Avanzini2021a} and introducing spatially organized compartments is straightforward. 
It suffices to treat the chemical species in the different compartments as different dynamical variables and add reactions amongst them to describe (passive or active) exchanges across compartments.
Adding diffusion by promoting the description of some or all species from homogeneous concentrations to space-dependent concentration fields is also in principle not an issue. In such cases diffusion is treated within Fick's law and contributes to the dissipation in the CRN~\cite{Falasco2018a, Avanzini2019a, Avanzini2020a}.  

Our approach is fundamentally different from flux balance analysis ones.
These latter are designed to determine the steady-state currents in a CRN. In the space of all possible steady-state currents, they select those that satisfy a set of constraints the system is supposedly subjected to. These can range from thermodynamic constraints~\cite{Gerstl2015,DeMartino2018,Niebel2019} or limits imposed by the environment~\cite{Akbari2021} to presumed aims like growth maximization for some cells~\cite{Niebel2019}.
The resulting steady-state currents will naturally depend on the enforced constraints.
These approaches are thus top-down. 
Some of the more teleological constraints, such as maximizing growth, may only be justified in complex systems such as living systems shaped by evolution. 
Identifying the constraints predicting the steady-state currents of the CRN in Fig.~\ref{fig:modularization} for a given set of thermodynamically consistent kinetic constants would be more complicated than solving the full dynamics.  
Instead, our circuit theory may be defined as bottom-up. Indeed, it is built to be compatible with a microscopic description of the dynamics. As explained in the main text, the current-concentration characteristics of the modules result from the full dynamics and can be used as an input to our theory to predict the correct dynamics and thermodynamics of the terminal species (Sec.~\ref{sec:circuit_description}). 

Our work shares some conceptual similarities with the work by Oster and coworkers which, in the seventies, developed a very general network thermodynamics describing networks made of any type of thermodynamics systems~\cite{Oster1971, Oster1973}. Their intent was to describe coupled thermodynamics processes arising in biophysics involving different forces such as mechanical, electrical, and chemical forces. 
Their theory makes use of bond graphs, a graphical representation inspired by electric diagrams used in electrical circuit theories. 
The generality of the theory turns however into a disadvantage in the context of CRNs, since the theory is not tailored for them.
The bond graph representation of simple CRNs quickly becomes very cumbersome~\cite{Oster1974, Perelson1974}. This also explains why the use of the theory has remained limited to simple CRNs. In contrast, our formalism is algebraic and based on the representation of modules in terms of emergent cycles. 
The latter identify the minimal set of currents needed to define the current-concentration characteristics of a module and to determine its dissipation.
They also provide an intuitive description of modules in terms of effective chemical reactions which can be easily represented in terms of hypergraphs.  
The theory by Oster and coworkers do not exploit that reduction. 
Furthermore, one of the main purposes of our theory is to provide a simplified (i.e., coarse-grained) description of the dynamics of CRNs. 
The theory by Oster and coworkers instead has been mostly used as a formalization and representation tool, not as a reduction tool. 

We now discuss interesting perspectives raised by our work.

Electronic engineering makes extensive use of circuit theory to design circuits with intended functionalities, such as computing operations.
Similarly, one should explore how to make use of the chemical circuit theory to design useful chemical functions. This may be particularly relevant in the context of chemical computing, a field increasingly raising attention~\cite{soloveichik2010, Chen2013, Nielsen2016Apr, winfree2019chemical}. 

Our work focused on the deterministic description of CRNs, but in many instances such as cellular biology, extending the theory to stochastic descriptions of CRNs would be important.
This may be challenging because the statistics of the effective chemical reactions is not trivially related to the Poisson statistics of the elementary reactions, see for instance Ref.~\cite{Sinitsyn2009}.

We presented our circuit theory for open CRNs exchanging matter and heat with the surrounding. But other forms of energy may be incorporated in the description, such as energy provided by thermal light~\cite{Penocchio2019a}, electrical energy, and osmotic pressure. Indeed, the concept of emergent cycles is a general feature of thermodynamics when taking into account conservation laws~\cite{Rao2018a}.  
This is why circuit theories have the potential to provide a powerful and realistic characterization of the dynamics and thermodynamics of complex systems.
The key point is that, as for electric circuits, the current-potential characteristics provide an empirical characterization of complex modules that would otherwise be very hard to determine. 

As shown implicitly in Ref.~\cite{Wachtel2022}, but clearly retrospectively, the circuit theory underlies the fact that central metabolism can be decomposed into modules (glycolysis, Krebs cycles,\dots). 
But what is true at the level of cellular metabolism still holds true at higher levels, namely, whenever one is dealing with open CRNs coupled to each other by the exchange of terminal species. A food web for instance can be seen as a collection of modules representing the metabolisms of the different living systems feeding on each other and ultimately powered by solar energy. In ecology, like previously in biochemistry, tracking the movement of different types of atoms across a network under different molecular forms is nowadays used to reconstruct CRNs up to global scales, as for instance in biogeochemistry~\cite{schlesinger2013,smith2016}. Measuring or estimating current-potential characteristics may not be easy in such a context, but is conceivable and worth trying given the importance of these networks. 

Circuit theories may even provide a proper framework to formulate models in ecological economics (also called steady-state economics) where minimizing the dissipation arising in the use and recycling of natural resources is a major concern~\cite{daly2004ecological}.

\begin{acknowledgments}
This research was supported by project ChemComplex (C21/MS/16356329) funded by FNR (Luxembourg), and by project INTER/FNRS/20/15074473 funded by F.R.S.-FNRS (Belgium) and FNR (Luxembourg).
\end{acknowledgments}

\appendix

\section{Formal Circuit Theory\label{app:def_chem_module}}
We start by summarizing the formal definition of CRNs on top of which we develop the concept of chemical modules.
We consider CRNs composed of chemical species, identified by the label $\chemspecies\in\setchemspecies$, 
undergoing elementary reactions~\cite{Svehla1993}, identified by the index $\elrct\in\setelrct$.
Each reaction $\elrct$ is represented by an equation like
\begin{equation}
\chemspeciesvec\cdot \stcoeff{+} \ch{<=>[ $+\elrct$ ][ $-\elrct$ ]}\chemspeciesvec\cdot \stcoeff{-}\,,
\label{eq:elementary_reaction}
\end{equation}
with $\chemspeciesvec =(\dots,\chemspecies,\dots)^\intercal_{\chemspecies\in\setchemspecies}$ the vector of chemical species and $\stcoeff{\pm}$ the vector of stoichiometric coefficients of the forward/backward reaction $\pm\elrct$.
Note that all reactions are assumed to be reversible for thermodynamic consistency.
In open CRNs (such as the CRN in Fig.~\ref{fig:modularization}a) the species $\setchemspecies$ are split into the \textit{internal} species~$\setinternal$ and the \textit{exchanged} species~$\setexchanged$. 
The former undergo \textit{only} the chemical reactions $\elrct\in\setelrct$. 
The latter undergo the chemical reactions $\rho \in \setelrct$ and they are \textit{also} exchanged with the environment the CRN is exposed to.

The state of deterministic CRNs is specified by the concentration vector $\conc=(\dots, [\chemspecies],\dots)^\intercal$. 
Its dynamics follows the rate equation
\begin{equation}
\dt \conc=\matS\, \curr + \excurr\,,
\label{eq:dynamics_crns}
\end{equation}
where we introduced the stoichiometric matrix $\matS$, the reaction current vector $ \curr$, and the exchange current vector $\excurr$.
The first term on the r.h.s. of Eq.~\eqref{eq:dynamics_crns}, i.e., $\matS \curr$, accounts for the variations to the concentrations due to the chemical reactions.
Each column ${\colS}_\elrct$ of $\matS$ is given by ${\colS}_\elrct =  \stcoeff{-} - \stcoeff{+} $.
The reaction current vector $ \curr=(\dots,\currel_\elrct,\dots)^\intercal_{\elrct\in\setelrct}$ specifies the net reaction current for every $\elrct$ reaction as the difference between the forward $\currel_{+\elrct}$ and backward reaction flux $\currel_{-\elrct}$:
\begin{equation}
\currel_{\elrct}=\currel_{+\elrct} - \currel_{-\elrct}\,.
\label{eq_net_current}
\end{equation}
In ideal CRNs, the fluxes $\currel_{\pm\elrct}$ of elementary reactions satisfy mass-action kinetics ~\cite{Groot1984, Pekar2005, Laidler1987}:
\begin{equation}
\currel_{\pm\elrct} = k_{\pm\elrct}\conc^{\stcoeff{\pm}}\,,
\label{eq_mass_action}
\end{equation}
where $k_{\pm\elrct}$ are the kinetic constants of the forward/backward reaction $\pm\elrct$ and we used the following notation $\boldsymbol a^{\boldsymbol b} = \prod_\muteindex a_\muteindex^{b_\muteindex}$.
The second term on the r.h.s. of Eq.~\eqref{eq:dynamics_crns}, i.e., $\excurr$, specifies the matter flows with the environment~\cite{Avanzini2022a}.
It has null entries for the internal species, i.e., $\excurrel_{\chemspecies}=0$ for $\chemspecies \in\setinternal$: the concentration of the internal species changes only because of the chemical reactions by definition.

By applying the splitting of the chemical species into internal and exchanged ones to the stoichiometric matrix
\begin{equation}
\matS=\begin{pmatrix}
\matSX \\ 
\matSY\\
\end{pmatrix}\,,
\end{equation}
and the concentration vector $\conc= (\intconc,\exconc)$, the rate equation~\eqref{eq:dynamics_crns} becomes
\begin{subequations}
\begin{align}
&\dt \intconc=\matSX \curr\label{eq:dynamics_crns_X}\,,\\
&\dt \exconc=\matSY \curr + \excurrY\label{eq:dynamics_crns_Y}\,,
\end{align}
\end{subequations}
with $ \excurrY = (\dots,\excurrel_\chemspecies,\dots )^\intercal_{\chemspecies\in\setexchanged}$ collecting the not null entries of $\excurr$. Note that the Eqs.~\eqref{eq:dynamics_crns_X} and~\eqref{eq:dynamics_crns_Y} are only a reformulation of Eq.~\eqref{eq:dynamics_crns}.

\subsection{Elementary Modules\label{app:def_chem_module_el}}
A chemical module of a CRN, labeled by the index $\module$, is defined as a subnetwork:
a subset of chemical reactions $\modsetelrct\subset\setelrct$ interconverting 
a subset of chemical species $\modsetchemspecies \subset\setchemspecies$.
The species $\modsetchemspecies$ are further classified as either internal species of the module~$\modsetinternal$ or terminal species~$\modsetexchanged$. 
The former \textit{must} only undergo the chemical reactions $\rho \in \modsetelrct$ (the reason for this will become clear at the end of App.~\ref{app:def_chem_module_eff}).
The latter undergo the chemical reactions $\rho \in \modsetelrct$, but they \textit{can} also undergo other reactions $\rho \notin \modsetelrct$ and/or be externally exchanged.
Several examples of modules are discussed in Sec.~\ref{sec:module_def} and illustrated in Fig.~\ref{fig:modularization}1.

The rate equation~\eqref{eq:dynamics_crns} can be specialized 
for the concentrations $\modintconc =(\dots,[\chemspecies],\dots){}^\intercal_{\chemspecies\in\modsetinternal}$ 
and for the concentrations $\modexconc =(\dots,[\chemspecies],\dots){}^\intercal_{\chemspecies\in\modsetexchanged}$: 
\begin{subequations}
\begin{align}
&\dt \modintconc= \matSQM\,\modcurr\,,\label{eq:dynamics_crns_Qm}\\
&\dt \modexconc=  \matSPM\,\modcurr + \modexcurr \,,\label{eq:dynamics_crns_Pm}
\end{align}
\end{subequations}
where $\modcurr=(\dots,\currel_\elrct,\dots){}^\intercal_{\elrct\in\modsetelrct}$, and we introduced the substoichiometric matrixes $\matSQM$ and $\matSPM$ whose entries are  $\{\elS^\chemspecies_\elrct\}^{\chemspecies\in\modsetinternal}_{\elrct\in\modsetelrct}$ and
$\{\elS^\chemspecies_\elrct\}^{\chemspecies\in\modsetexchanged}_{\elrct\in\modsetelrct}$, respectively.
Here, the terminal current vector of the module $\modexcurr$ accounts for all the processes affecting the concentrations $\modexconc$ besides the reactions $\rho \in \modsetelrct$. 
Note that Eq.~\eqref{eq:dynamics_crns_Qm} and~\eqref{eq:dynamics_crns_Pm} coincide with Eq.~\eqref{eq:dynamics_crns_X} and~\eqref{eq:dynamics_crns_Y} when the module is treated as an open CRN,
and the species $\modsetinternal$ and $\modsetexchanged$ are identified as $\setinternal$ and $\setexchanged$, respectively.

\subsection{Effective Modules at Quasi Steady State\label{app:def_chem_module_eff}}
Modules can be coarse grained into effective reactions interconverting the terminal species~$\modsetexchanged$.
This can be done when two conditions are satisfied. 
The first is the existence, for every concentration vector $\modexconc$, of a unique steady-state concentration vector $\modintconcpss (\modexconc)$ for the internal species of the module (see App.~\ref{sub:instability} for an explicit example where this does not hold, but the theory can still be applied).
The second is the equivalence between the actual concentration vector $\modintconc$  and the steady-state one $\modintconcpss$.
This obviously happens at the steady state to which the module relaxes when the concentrations $\modexconc$ are kept constant by the other reactions $\elrct\notin\modsetelrct$ and the exchange processes, i.e., when
\begin{equation}
\modexcurr = - \matSPM\,\modcurr\,.
\label{eq:module_excurrent_fixing_conc}
\end{equation}
It also happens to a very good approximation when the chemical species evolve over two different time scales such that the concentrations of the $\modsetinternal$ species quickly relax to the steady state corresponding to the values of $\modexconc$.
Indeed, a zero-order expansion of the concentrations $\modintconc$ in the ratio between the fast time scale of the internal species and the slow time scale of the terminal species leads to
\begin{equation}
\modintconc = \modintconcpss (\modexconc)\,\,\,\forall t\,.
\label{eq:qss_assumption}
\end{equation}
This physically occurs when i) the elementary reactions and the exchange processes involving only the terminal species are slower than the elementary reactions involving only the internal species and ii) the abundance of the terminal species is very large compared to the abundance of the internal species which therefore changes much more quickly~\cite{Sinitsyn2009} (indeed,
when the terminal and internal species are involved in the same reaction, 
on the same time scale the concentrations of the internal species dramatically change, the concentrations of the terminal species remain almost constant).
Note that describing electronic components in terms of their I-V curves also requires a time-scale separation between their internal dynamics and the dynamics of the voltages on their contacts or pins.

When those two conditions are satisfied, the reaction current vector of the module 
\begin{equation}
\modeffcurr\equiv\modcurr( \modintconcpss,\modexconc)
\label{eq:quasi_ss_current_definition}
\end{equation}
depends only on the concentrations $\modexconc$ and 
is, by definition, a steady-state current of Eq.~\eqref{eq:dynamics_crns_Qm}, namely, $\matSQM \modeffcurr=0 $.
This means that $\modeffcurr\in\mathrm{ker}(\matSQM)$ and, consequently, 
can be written as 
\begin{equation}
\modeffcurr=\sum_{\modcycleindex}\modcycle_\modcycleindex\,\modcyclecurrel_\modcycleindex\,,
\label{eq:quasi_ss_current_decomposition}
\end{equation}
using the (linearly independent) right-null vectors $\matSQM\modcycle_\modcycleindex=0$ and $\modexconc$ dependent coefficients $\{\modcyclecurrel_\modcycleindex\}$.
The vectors $\{\modcycle_\modcycleindex\}$ are called cycles because they represent sequences of reactions that upon completion leave the concentrations $\modintconc$ unchanged.
Each coefficient $\modcyclecurrel_\modcycleindex$ represents the current along the cycle $\modcycleindex$.
The cycles can be split into two disjoint sets, i.e., $\{\modcycle_\modcycleindex\}= \{\modcycle_\modintcycleindex \}\cup\{\modcycle_\modemcycleindex\}$. 
The so-called \textit{internal} cycles $\{\modcycle_\modintcycleindex \}$ are also right-null vectors of $\matSPM$, i.e., $\matSPM \modcycle_\modintcycleindex= 0$.
They thus represent sequences of reactions that upon completion leave also the concentrations $\modexconc$ unchanged (for instance, the (aqua green) module~$\lmd$ in Fig.~\ref{fig:modularization}a has two internal cycles as specified in Fig.~\ref{fig:modularization_stoichiometry}).
The others $\{\modcycle_\modemcycleindex\}$ are called \textit{emergent} cycles.
The internal and emergent cycles of the modules in Fig.~\ref{fig:modularization}a are reported in Fig.~\ref{fig:modularization_stoichiometry}.

By employing the steady-state current~\eqref{eq:quasi_ss_current_decomposition} in Eq.~\eqref{eq:dynamics_crns_Pm} and the splitting of the cycles into internal and emergent ones, we obtain an effective and closed dynamical equation for the $\modsetexchanged$ species
\begin{subequations}
\begin{align}
\dt \modexconc
&=  \matSPM\,\modeffcurr + \modexcurr\\
&= \sum_{\modemcycleindex}\matSPM\,\modcycle_\modemcycleindex\,\modcyclecurrel_\modemcycleindex + \modexcurr\,.
\label{eq:eff_rate_eq_module_v1}
\end{align}
\end{subequations}
Each vector $\matSPM\,\modcycle_\modemcycleindex$ specifies the net variation of the number of molecules for each $\modsetexchanged$ species along the $\modemcycleindex$ emergent cycle. 
Namely, it specifies the stoichiometry of an effective reaction.
The effective reactions of the modules in Fig.~\ref{fig:modularization}a are discussed in Sec.~\ref{sec:module_def}.
Correspondingly, the emergent cycle current $\modcyclecurrel_\modemcycleindex$ specifies the current of this effective reaction.

Equation~\eqref{eq:eff_rate_eq_module_v1} can be rewritten in a more compact way,
\begin{equation}
\dt \modexconc = \effmatSPM\modeffcurrcycle + \modexcurr
\label{eq:eff_rate_eq_module}
\end{equation}
introducing the effective stoichiometric matrix $\effmatSPM$ and the effective current vector $\modeffcurrcycle$.
Here, each $\modemcycleindex$ column of $\effmatSPM$ is given by $\matSPM\,\modcycle_\modemcycleindex$ and $\modeffcurrcycle =(\dots, \modcyclecurrel_\modemcycleindex,\dots)^\intercal$  does not, in general, satisfy mass-action kinetics.

Note that each module is defined as a subnetwork with a unique set of internal species because this ensures that $\modintconcpss$ and, consequently, $\modeffcurrcycle$ are functions of the concentrations $\modexconc$ of its terminal species only.
This is a necessary condition to obtain the closed dynamical equation~\eqref{eq:eff_rate_eq_module} for the terminal species.

\subsection{Effective Modules with Internal Oscillations\label{app:def_chem_module_eff_oscillations}} 
Modules can be coarse grained into the same effective reactions defined by the emergent cycles $\{\modcycle_\modemcycleindex\}$ even if the concentrations of the internal species oscillate.
This can be done when the concentration vector $\modintconc$ relaxes instantaneously to an oscillating dynamics $\modintconcosc(\modexconc, t)$ (formally, we use again a zero-order expansion of $\modintconc$ in the ratio between the relaxation time scale of the internal species and the time scale of the terminal species leading to $\modintconc = \modintconcosc(\modexconc, t)$), whose period~$\period$ is much shorter than the time scale $\slowt$ of the terminal species, i.e., $\period/\slowt\ll1$.

When these conditions are satisfied, 
the reaction current vector of the module 
\begin{equation}
\modosccurr\equiv\modcurr( \modintconcosc,\modexconc)
\end{equation}
is not a steady-state current of Eq.~\eqref{eq:dynamics_crns_Qm} and it reads 
\begin{equation}
\modosccurr=
\sum_{\modintcycleindex}\modcycle_\modintcycleindex\,\modcyclecurrel_\modintcycleindex + 
\sum_{\modemcycleindex}\modcycle_\modemcycleindex\,\modcyclecurrel_\modemcycleindex +  
\sum_{\modcocycleindex}\modcocycle_\modcocycleindex\,\modcyclecurrel_\modcocycleindex \,,
\end{equation}
where we used the internal $\{\modcycle_\modintcycleindex\}$ and emergent $\{\modcycle_\modemcycleindex\}$ cycles as well as the (linearly independent) vectors $\{\modcocycle_\modcocycleindex\}$, named cocycles, generating the orthogonal complement of $\mathrm{ker}(\matSQM)$.
Correspondingly, the dynamical equation~\eqref{eq:dynamics_crns_Pm} for the $\modsetexchanged$ species becomes 
{\small\begin{equation}
\dt \modexconc = 
{\matSPM}\int_t^{t+\slowt}\frac{\de t'}{\slowt}
\Big[\sum_{\modemcycleindex}\modcycle_\modemcycleindex\,\modcyclecurrel_\modemcycleindex+ 
\sum_{\modcocycleindex}\modcocycle_\modcocycleindex\,\modcyclecurrel_\modcocycleindex \Big] + 
\modexcurr
\label{eq:dynamics_crns_Pm_oscillation}
\end{equation}}
by using $\matSPM\modcycle_\modintcycleindex = 0$ and assuming that the concentrations of all terminal species are almost constant in the time interval $\slowt$, namely, $\dt \modexconc \simeq (\modexconc(t+\slowt) - \modexconc(t))/\slowt $ and $\int_t^{t+\slowt}\de t'\text{ }\modexcurr/\slowt \simeq \modexcurr$. 

We now show that Eq.~\eqref{eq:dynamics_crns_Pm_oscillation} simplifies to a closed dynamical equation for the $\modsetexchanged$ species similar to Eq.~\eqref{eq:eff_rate_eq_module}.
To do so, we consider that 
i) the internal dynamics completes $\nosc$ oscillations (with $\slowt/\period - 1< \nosc\leq \slowt/\period)$ in the time interval $\slowt$;
ii) $\nosc \simeq \slowt/\period $ when the time scale separation is satisfied, i.e., $\period/\slowt\ll1$;
iii) the integral $\int_t^{t+\slowt}\de t'/\slowt$ can be split into $\int_t^{t+\nosc\period}\de t'/\slowt + \int_{t+\nosc\period}^{t+\slowt}\de t'/\slowt$, where the latter contribution is of order $1-\nosc\period/\slowt \simeq 0$ and hence negligible;
iv) $\int_t^{t+\nosc\period}\de t' \modcyclecurrel_\modcocycleindex = 0 $ since $\int_t^{t+\period}\de t'\text{ }\modosccurr\in \mathrm{ker}(\matSQM)$.
These lead to 
\begin{equation}
\dt \modexconc = 
 \sum_{\modemcycleindex}\matSPM\modcycle_\modemcycleindex\,\osc{\modcyclecurrel}_\modemcycleindex+ 
\modexcurr\,,
\label{eq:eff_rate_eq_module_v2}
\end{equation}
where $\osc{\modcyclecurrel}_\modemcycleindex \equiv \int_t^{t+\period}\de t'{\modcyclecurrel}_\modemcycleindex/\period $ is a function of $\modexconc$ only.
Similarly to Eq.~\eqref{eq:eff_rate_eq_module_v1}, also Eq.~\eqref{eq:eff_rate_eq_module_v2} can be rewritten as Eq.~\eqref{eq:eff_rate_eq_module} by using the effective stoichiometric matrix $\effmatSPM$ and collecting $\{\osc{\modcyclecurrel}_\modemcycleindex\}$ into an effective current vector.
This physically means that on the time scale $\slowt$ of the terminal species, the internal dynamics is averaged over many ($\sim\slowt/\period\gg1$) oscillations and acts, in practice, as an effective steady state.

\subsection{Effective Currents via the Elementary Mechanism\label{app:definition_current_analytical}}
We show here how to determine the effective currents, i.e., the function  $\modeffcurrcycle(\modexconc)$ in Eq.~\eqref{eq:eff_rate_eq_module}, 
from the elementary dynamics, given in Eq.~\eqref{eq:dynamics_crns_Qm} and~\eqref{eq:dynamics_crns_Pm},
by assuming that the steady-state concentration vector $\modintconcpss (\modexconc)$ can be computed for every $\modexconc$ (either analytically or numerically). 
This approach is then illustrated in App.~\ref{app:illustration_current_analytical} for the (blue) module~$\lma$, the (green) module~$\lmb$ and the (purple) module~$\lme$ of Fig.~\ref{fig:modularization}a.

We start by recognizing that $\modeffcurr(\modexconc)$ can be obtained using its definition~\eqref{eq:quasi_ss_current_definition} and $\modintconcpss (\modexconc)$.
We then rewrite Eq.~\eqref{eq:quasi_ss_current_decomposition} as 
\begin{equation}
\modeffcurr(\modexconc) = \matCM \,\modcurrcycle(\modexconc)\,,
\end{equation}
where we introduced the cycle current vector $\modcurrcycle(\modexconc) = (\dots,\modcyclecurrel_\modcycleindex(\modexconc),\dots)^\intercal$, which  includes also the internal cycle currents unlike $\modeffcurrcycle(\modexconc)$ in Eq.~\eqref{eq:eff_rate_eq_module} which includes only the emergent cycles, and 
the matrix $\matCM$ whose columns are the cycles $\{\modcycle_\modcycleindex\}$.
Since the cycles $\{\modcycle_\modcycleindex\}$ are linear independent, the matrix $\matCM^\intercal \matCM$ can be inverted, and we thus obtain
\begin{equation}
\modcurrcycle(\modexconc) =  (\matCM^\intercal \matCM)^{-1}\matCM^\intercal \,\modeffcurr(\modexconc) \,.
\label{eq:cyclecurr1}
\end{equation}

\subsection{Effective Currents via the Terminal Currents\label{app:definition_current_experimental}}
We now discuss how to determine the effective currents, i.e., the function of  $\modeffcurrcycle(\modexconc)$ in Eq.~\eqref{eq:eff_rate_eq_module}, by assuming that
i) the effective stoichiometric matrix $\effmatSPM$ is known and 
ii) the concentrations $\modexconc$ can be kept equal to arbitrary and constant values by controlling the terminal currents $\modexcurr$ according to  Eq.~\eqref{eq:module_excurrent_fixing_conc}.
This approach is illustrated in Sec.~\ref{sec:currents} of the main text.

When the concentrations $\modexconc$ are constant, the module relaxes instantaneously towards a nonequilibrium steady state.
By using Eq.~\eqref{eq:eff_rate_eq_module}, the steady-state terminal currents of the module read 
\begin{equation}
\modssexcurr(\modexconc) = -\effmatSPM \modeffcurrcycle(\modexconc)\,.
\label{eq:module_ss_exucrrent}
\end{equation}
We now use Eq.~\eqref{eq:module_ss_exucrrent} to express $\modeffcurrcycle(\modexconc)$ in terms of $\modssexcurr(\modexconc)$.
To do so, we recognize that the effective stoichiometric matrix $\effmatSPM$ has no right-null vectors, as already discussed in Ref.~\cite{Avanzini2020b}.
Indeed, suppose that there is a vector $\vect=(\dots,\vectel_\modemcycleindex,\dots)^\intercal$ such that $\effmatSPM\vect = 0$.
This means that 
\begin{equation}
\matSPM\sum_{\modemcycleindex}\modcycle_\modemcycleindex \vectel_\modemcycleindex =0\,,
\end{equation}
and, consequently, $\sum_{\modemcycleindex}\modcycle_\modemcycleindex \vectel_\modemcycleindex$ is a right-null vector of both $\matSQM$ and $\matSPM$, i.e., an internal cycle.
Since $\sum_{\modemcycleindex}\modcycle_\modemcycleindex \vectel_\modemcycleindex$ is a linear combination of emergent cycles, we can conclude that $\effmatSPM$ has no right-null vectors.
This implies that the columns of $\effmatSPM$ are linearly independent and the matrix $\big(\effmatSPM\big)^\intercal\effmatSPM$ can be inverted. Thus,
\begin{equation}
 \modeffcurrcycle(\modexconc) = - \big(\big(\effmatSPM\big)^\intercal\effmatSPM\big)^{-1}\big(\effmatSPM\big)^\intercal \, \modssexcurr(\modexconc) \,.
\label{eq:cyclecurr2}
\end{equation}

\subsection{General Circuit Theory and Kirchhoff’s Laws\label{app:kirchhoff}}
Once modules are fully characterized, namely, their effective reactions $\{\matSPM\,\modcycle_\modemcycleindex\}$ and currents $\{\modeffcurrcycle\}$ are known, they can be connected by sharing the terminal species.
The result is a circuit theory where the dynamics of all terminal species emerges from combining the dynamical equation~\eqref{eq:eff_rate_eq_module} of the modules:
\begin{equation}
\dt \terconc = \matSeff {\cyclecurr} + \excurr\,.
\label{eq:rate_eq_ct_app}  
\end{equation}
Here, $\terconc = (\dots,[\chemspecies],\dots)^\intercal_{\chemspecies\in\setterminal}$ is the concentration vector of all terminal species (with $\setterminal$ the set of all terminal species $\setterminal =\bigcup_\module \modsetexchanged$),
$\matSeff$ (resp. $\cyclecurr$) is the stoichiometric matrix (resp. effective current vector) whose columns (resp. entries) specify the net stoichiometry (resp. current) of the effective reactions of all modules (labeled $\emer$),
and $\excurr$ is the exchange current vector as in Eq.~\eqref{eq:dynamics_crns}.
Note that Eq.~\eqref{eq:rate_eq_ct_app} is exactly equivalent to Eq.~\eqref{eq:rate_eq_ct} in the main text.

Equation~\eqref{eq:rate_eq_ct_app} satisfies the chemical equivalent of Kirchhoff’s laws in electrical circuits (whose specific expressions for the CRN in Fig.~\ref{fig:modularization}b are reported in Sec.~\ref{sub:kirchhoff}).
The chemical equivalent of \textit{Kirchhoff’s current law} is the condition that the sum of all the currents (both effective and exchange) affecting the concentration of each terminal species vanishes at steady state (denoted by the overline):
\begin{equation}
\dt \ssterconc = \matSeff {\sscyclecurr} + \ssexcurr = 0\,.
\end{equation}
The chemical equivalent of \textit{Kirchhoff’s voltage law} is the condition that the sum of the variations of the Gibbs free energy along each internal cycle of $\matSeff$ vanishes:
\begin{equation}
    \vfreerct \cdot \cyclect_{\inter} = 0\,,
\end{equation}
where $\vfreerct = (\dots,\freerct{\emer},\dots)^\intercal$ (with $\freerct{\emer}$ the variations of the Gibbs free energy along the effective reaction $\emer$, i.e., $\freerct{\emer}=\sum_{\chemspecies\in\setterminal}\chempotential_\chemspecies \elSeff^\chemspecies_\emer
$) and the internal cycles are defined as $\matSeff\cyclect_{\inter} = 0$ (note that this definition is analogous to the one given for the internal cycles of modules in App.~\ref{app:def_chem_module_eff}).


\section{Illustration of the Analytical and Computational Derivation of the Effective Currents\label{app:illustration_current_analytical}}
We illustrate here two approaches to determine the current-concentration characteristic of some of the modules in Fig.~\ref{fig:modularization}a when their elementary dynamics is known and can be solved (namely, when steady-state concentration vector $\modintconcpss (\modexconc)$ can be computed as discussed in general in App.~\ref{app:definition_current_analytical}).

\subsection{Analytical Strategy~\label{sub:effcurr_analytical}}
Explicit analytical expressions can be derived for the current-concentration characteristic when the internal reactions of a module are pseudo-first-order reactions, i.e. when they are effectively unimolecular reactions in terms of the internal species, and follow mass-action kinetics~\cite{Groot1984,Laidler1987,Pekar2005} (see also App.~\ref{app:def_chem_module}) with known kinetic constants.
To do so, the diagrammatic method developed in Ref.~\cite{King1956, Hill1966} can be used, as done in Refs.~\cite{Gunawardena2012, Wachtel2018}.
This strategy can be applied for all the modules in Fig.~\ref{fig:modularization}a except the (purple) module~$\lme$ (see Subs.~\ref{sub:effcurr_numerical}). 
We now show it for the (blue) module~$\lma$ and the (green) module~$\lmb$.

Let us start with the (blue) module~$\lma$ in Fig.~\ref{fig:modularization}a.
The diagrammatic method~\cite{King1956, Hill1966} provides the steady-state concentrations of the internal species of the module $(\steady{[\ch{E_{$\lma$}}]}, \steady{[\ch{E_{$\lma$}S}]}, \steady{[\ch{E_{$\lma$}S_2}]})$ for given concentrations of the terminal species $([\ch{S}], [\ch{N_{$\lma$}}])$:
\begin{subequations}
\begin{align}
&\steady{[\ch{E_{$\lma$}}]}=\frac{\consquantity_{\ch{E}_\lma}}{\mathcal D_{\lma}}\big( k_{-\lma_{1}}k_{-\lma_{3}} + k_{+\lma_{2}}k_{-\lma_{3}} \big) \,,\\
&\steady{[\ch{E_{$\lma$}S}]}=\frac{\consquantity_{\ch{E}_\lma}}{\mathcal D_{\lma}}\big( k_{+\lma_{1}}k_{-\lma_{3}} [\ch{S}]+ k_{-\lma_{2}}k_{-\lma_{3}} [\ch{N}_\lma] \big)\,,\\
&\steady{[\ch{E_{$\lma$}S_2}]}=\frac{\consquantity_{\ch{E}_\lma}}{\mathcal D_{\lma}}\big( k_{+\lma_{1}}k_{+\lma_{3}} [\ch{S}]^2+ k_{-\lma_{2}}k_{+\lma_{3}}  [\ch{S}][\ch{N}_\lma] \big)\,,
\end{align}
\end{subequations}
where $\{k_{\pm\lma_\muteindex}\}_{\muteindex=1,2,3}$ are the kinetic constants of the chemical reactions in~\eqref{eq:module1_rct}, $\consquantity_{\ch{E}_\lma} =  {[\ch{E_{$\lma$}}]} + {[\ch{E_{$\lma$}S}]} + {[\ch{E_{$\lma$}S_2}]}$ is the total concentration of the enzyme which is conserved by the chemical reactions and 
{\small\begin{equation}
\mathcal D_{\lma} = 
(k_{-\lma_{1}} + k_{+\lma_{2}})k_{-\lma_{3}} +
(k_{+\lma_{3}} [\ch{S}] + k_{-\lma_{3}})(k_{+\lma_{1}}[\ch{S}] + k_{-\lma_{2}} [\ch{N}_\lma]) \,.
\end{equation}}
Thus, the steady-state reaction currents of the internal reactions~\eqref{eq:module1_rct}, according to mass-action kinetics, are specified as
\begin{subequations}
{\small
\begin{align}
&\smodsscurrel{\lma_1}  =\frac{\consquantity_{\ch{E}_\lma}}{\mathcal D_{\lma}}  \big(
k_{+\lma_{1}}k_{+\lma_{2}}k_{-\lma_{3}} [\ch{S}]+ k_{-\lma_{1}}k_{-\lma_{2}}k_{-\lma_{3}} [\ch{N}_\lma]  \big) \,, \\
&\smodsscurrel{\lma_2} = \smodsscurrel{\lma_1}\,,\\
&\smodsscurrel{\lma_3} =0\,.
\end{align}}
\end{subequations}
Since the current vector $\smodsscurr{\lma} = (\smodsscurrel{\lma_1}, \smodsscurrel{\lma_2}, \smodsscurrel{\lma_3})^\intercal$ must be equal to the emergent cycle~\eqref{eq:module1_emcycle} times the effective reaction current $\smodcyclecurrel{\emer_\lma}$~(refer to App.~\ref{app:def_chem_module_eff} for a formal discussion), i.e., 
\begin{equation}
\smodsscurr{\lma} = \smodcyclecurrel{\emer_\lma} \kbordermatrix{
   	&		\cr
      	&1		\cr
     	&1  		\cr
    	&0  		\cr
  }\,,
\end{equation}
we obtain an analytical expression of the effective reaction current:
\begin{equation}
\smodcyclecurrel{\emer_\lma}
 =\frac{\consquantity_{\ch{E}_\lma}}{\mathcal D_{\lma}}  \big(
k_{+\lma_{1}}k_{+\lma_{2}}k_{-\lma_{3}} [\ch{S}]+ k_{-\lma_{1}}k_{-\lma_{2}}k_{-\lma_{3}} [\ch{N}_\lma]  \big) \,.
\label{eq:module1_emcyclecurr}
\end{equation}

We now turn to the (green) module~$\lmb$ in Fig.~\ref{fig:modularization}a.
Using again the diagrammatic method~\cite{King1956, Hill1966}, 
we determine the steady-state concentrations of the internal species of the module $(\steady{[\ch{E_{$\lmb$}}]}, \steady{[\ch{E_{$\lmb$}F}]}, \steady{[\ch{E_{$\lmb$}W}]}, \steady{[\ch{E^{*}_{$\lmb$}}]})$
for fixed concentrations of the terminal species $([\ch{N_{$\lma$}}],[\ch{N_{$\lmb$}}],[\ch{F}],[\ch{W}])$:
{\small\begin{equation}
\begin{split}
\steady{[\ch{E_{$\lmb$}}]}=\frac{\consquantity_{\ch{E}_\lmb}}{\mathcal D_{\lmb}}\big(
& k_{-\lmb_{1}}k_{-\lmb_{4}}k_{-\lmb_{5}}[\ch{N_{$\lmb$}}] + k_{-\lmb_{1}}k_{-\lmb_{2}}k_{-\lmb_{4}} \\
+ & k_{-\lmb_{1}}k_{-\lmb_{2}}k_{+\lmb_{5}} + k_{-\lmb_{1}}k_{+\lmb_{3}}k_{-\lmb_{4}} \\
+ & k_{-\lmb_{1}}k_{+\lmb_{3}}k_{+\lmb_{5}} + k_{+\lmb_{2}}k_{+\lmb_{3}}k_{+\lmb_{5}} \\
+ & k_{+\lmb_{2}}k_{+\lmb_{3}}k_{-\lmb_{4}} + k_{+\lmb_{3}}k_{+\lmb_{4}}k_{+\lmb_{5}}[\ch{N_{$\lma$}}]  \big)
\end{split} 
\label{eq:model2_ssconc1}
\end{equation}}
{\small\begin{equation}
\begin{split}
\steady{[\ch{E_{$\lmb$}F}]}=\frac{\consquantity_{\ch{E}_\lmb}}{\mathcal D_{\lmb}}\big(
& k_{+\lmb_{1}}k_{-\lmb_{4}}k_{-\lmb_{5}}[\ch{F}][\ch{N_{$\lmb$}}] + k_{+\lmb_{1}}k_{-\lmb_{2}}k_{-\lmb_{4}} [\ch{F}]\\
+ & k_{+\lmb_{1}}k_{-\lmb_{2}}k_{+\lmb_{5}}[\ch{F}] + k_{+\lmb_{1}}k_{+\lmb_{3}}k_{-\lmb_{4}}[\ch{F}] \\
+ & k_{+\lmb_{1}}k_{+\lmb_{3}}k_{+\lmb_{5}} [\ch{F}]+ k_{-\lmb_{2}}k_{-\lmb_{3}}k_{+\lmb_{5}}[\ch{W}] \\
+ & k_{-\lmb_{2}}k_{-\lmb_{3}}k_{-\lmb_{4}}[\ch{W}] + k_{-\lmb_{3}}k_{-\lmb_{4}}k_{-\lmb_{5}} [\ch{W}][\ch{N_{$\lmb$}}] \big)
\end{split} 
\label{eq:model2_ssconc2}
\end{equation}}
{\small\begin{equation}
\begin{split}
\steady{[\ch{E_{$\lmb$}W}]}=\frac{\consquantity_{\ch{E}_\lmb}}{\mathcal D_{\lmb}}\big(
& k_{+\lmb_{1}}k_{+\lmb_{4}}k_{+\lmb_{5}}[\ch{F}][\ch{N_{$\lma$}}] + k_{+\lmb_{1}}k_{+\lmb_{2}}k_{-\lmb_{4}}[\ch{F}] \\
+ & k_{+\lmb_{1}}k_{+\lmb_{2}}k_{+\lmb_{5}} [\ch{F}] + k_{-\lmb_{1}}k_{-\lmb_{3}}k_{-\lmb_{4}}[\ch{W}] \\
+ & k_{-\lmb_{1}}k_{-\lmb_{3}}k_{+\lmb_{5}}[\ch{W}] + k_{+\lmb_{2}}k_{-\lmb_{3}}k_{+\lmb_{5}}[\ch{W}] \\
+ & k_{+\lmb_{2}}k_{-\lmb_{3}}k_{-\lmb_{4}}[\ch{W}] + k_{-\lmb_{3}}k_{+\lmb_{4}}k_{+\lmb_{5}}[\ch{W}] [\ch{N_{$\lma$}}]  \big)
\end{split} 
\label{eq:model2_ssconc3}
\end{equation}}
{\small\begin{equation}
\begin{split}
\steady{[\ch{E_{$\lmb$}^{*}}]}=\frac{\consquantity_{\ch{E}_\lmb}}{\mathcal D_{\lmb}}\big(
& k_{+\lmb_{1}}k_{+\lmb_{4}}k_{-\lmb_{5}}[\ch{F}][\ch{N_{$\lma$}}][\ch{N_{$\lmb$}}]+ k_{+\lmb_{1}}k_{-\lmb_{2}}k_{+\lmb_{4}}[\ch{F}][\ch{N_{$\lma$}}] \\
+ & k_{+\lmb_{1}}k_{+\lmb_{2}}k_{-\lmb_{5}}[\ch{F}][\ch{N_{$\lmb$}}] + k_{+\lmb_{1}}k_{+\lmb_{3}}k_{+\lmb_{4}}[\ch{F}][\ch{N_{$\lma$}}]  \\
+ & k_{-\lmb_{1}}k_{-\lmb_{3}}k_{-\lmb_{5}}[\ch{W}][\ch{N_{$\lmb$}}] + k_{+\lmb_{2}}k_{-\lmb_{3}}k_{-\lmb_{5}}[\ch{W}][\ch{N_{$\lmb$}}] \\
+ & k_{-\lmb_{2}}k_{-\lmb_{3}}k_{+\lmb_{4}}[\ch{W}][\ch{N_{$\lma$}}]  + k_{-\lmb_{3}}k_{+\lmb_{4}}k_{-\lmb_{5}} [\ch{W}][\ch{N_{$\lma$}}][\ch{N_{$\lmb$}}] \big)
\end{split} 
\label{eq:model2_ssconc4}
\end{equation}}
where $\consquantity_{\ch{E}_\lmb} = {[\ch{E_{$\lmb$}}]} + {[\ch{E_{$\lmb$}F}]} + {[\ch{E_{$\lmb$}W}]} + {[\ch{E_{$\lmb$}^{*}}]}$ and $\mathcal D_{\lmb}$ equals the sum of all terms between parentheses in Eqs.~\eqref{eq:model2_ssconc1},~\eqref{eq:model2_ssconc2},~\eqref{eq:model2_ssconc3}, and~\eqref{eq:model2_ssconc4}.
The steady-state reaction currents of reactions $\lmb_1$, $\lmb_2$, $\lmb_3$, $\lmb_4$, and $\lmb_5$. 
 can thus be computed using again mass-action kinetics:
{\small\begin{equation}
\begin{split}
\smodsscurrel{\lmb_1} 
= &\frac{\consquantity_{\ch{E}_\lmb}}{\mathcal D_{\lmb}}\Big( k_{+\lmb_{1}}k_{+\lmb_{3}}\big(k_{+\lmb_{2}}( k_{-\lmb_{4}} + k_{+\lmb_{5}}) + k_{+\lmb_{4}}k_{+\lmb_{5}}[\ch{N_{$\lma$}}] \big)[\ch{F}]\\
- & k_{-\lmb_{1}}k_{-\lmb_{3}}\big(k_{-\lmb_{2}}(k_{-\lmb_{4}} + k_{+\lmb_{5}}) + k_{-\lmb_{4}}k_{-\lmb_{5}}[\ch{N_{$\lmb$}}]\big)[\ch{W}] \Big)
\end{split}
\end{equation}}
{\small\begin{equation}
\begin{split}
\smodsscurrel{\lmb_2} 
= &\frac{\consquantity_{\ch{E}_\lmb}}{\mathcal D_{\lmb}}\Big( k_{+\lmb_{1}}k_{+\lmb_{2}} \big(k_{+\lmb_{3}}(k_{-\lmb_{4}}+k_{+\lmb_{5}})+ k_{-\lmb_{4}}k_{-\lmb_{5}} [\ch{N_{$\lmb$}}] \big)[\ch{F}] \\
+ & k_{+\lmb_{2}}k_{-\lmb_{3}}k_{-\lmb_{4}}k_{-\lmb_{5}} [\ch{W}][\ch{N_{$\lmb$}}] - k_{+\lmb_{1}}k_{-\lmb_{2}}k_{+\lmb_{4}}k_{+\lmb_{5}}[\ch{F}][\ch{N_{$\lma$}}] \\
- & k_{-\lmb_{2}}k_{-\lmb_{3}}\big(k_{-\lmb_{1}}(k_{-\lmb_{4}}+k_{+\lmb_{5}})+k_{+\lmb_{4}}k_{+\lmb_{5}}[\ch{N_{$\lma$}}] \big)[\ch{W}]\Big)
\end{split}
\end{equation}}
{\small\begin{equation}
\begin{split}
\smodsscurrel{\lmb_3} 
= & \frac{\consquantity_{\ch{E}_\lmb}}{\mathcal D_{\lmb}} \Big(k_{+\lmb_{1}}k_{+\lmb_{3}}\big(k_{+\lmb_{2}}( k_{-\lmb_{4}} + k_{+\lmb_{5}}) + k_{+\lmb_{4}}k_{+\lmb_{5}}[\ch{N_{$\lma$}}] \big)[\ch{F}]\\
- & k_{-\lmb_{1}}k_{-\lmb_{3}}\big(k_{-\lmb_{2}}(k_{-\lmb_{4}} + k_{+\lmb_{5}}) + k_{-\lmb_{4}}k_{-\lmb_{5}}[\ch{N_{$\lmb$}}]\big)[\ch{W}] \Big)
\end{split}
\end{equation}}
{\small\begin{equation}
\begin{split}
\smodsscurrel{\lmb_4} 
= &\frac{\consquantity_{\ch{E}_\lmb}}{\mathcal D_{\lmb}}\Big( k_{+\lmb_{4}}k_{+\lmb_{5}} \big(k_{+\lmb_{1}}(k_{-\lmb_{2}} + k_{+\lmb_{3}}) [\ch{F}] + k_{-\lmb_{2}}k_{-\lmb_{3}}  [\ch{W}]\big)[\ch{N_{$\lma$}}] \\
- & k_{-\lmb_{4}}k_{-\lmb_{5}}\big(k_{-\lmb_{3}}(k_{-\lmb_{1}} + k_{+\lmb_{2}})[\ch{W}] + k_{+\lmb_{1}}k_{+\lmb_{2}}[\ch{F}]\big)[\ch{N_{$\lmb$}}]\Big)
\end{split}
\end{equation}}
{\small\begin{equation}
\begin{split}
\smodsscurrel{\lmb_5} 
= & \frac{\consquantity_{\ch{E}_\lmb}}{\mathcal D_{\lmb}}\Big(k_{+\lmb_{4}}k_{+\lmb_{5}} \big(k_{+\lmb_{1}}(k_{-\lmb_{2}} + k_{+\lmb_{3}}) [\ch{F}] + k_{-\lmb_{2}}k_{-\lmb_{3}}  [\ch{W}]\big)[\ch{N_{$\lma$}}] \\
- & k_{-\lmb_{4}}k_{-\lmb_{5}}\big(k_{-\lmb_{3}}(k_{-\lmb_{1}} + k_{+\lmb_{2}})[\ch{W}] + k_{+\lmb_{1}}k_{+\lmb_{2}}[\ch{F}]\big)[\ch{N_{$\lmb$}}]\Big)\,.
\end{split}
\end{equation}}
The corresponding current vector $\smodsscurr{\lmb} = (\smodsscurrel{\lmb_1},\smodsscurrel{\lmb_2},\smodsscurrel{\lmb_3},\smodsscurrel{\lmb_4},\smodsscurrel{\lmb_5})^\intercal$ can be written as a linear combination of the two emergent cycles ${\cycle_{\emer_\lmb}}$ and ${\cycle_{\emer_\lmb'}}$ in Fig.~\ref{fig:modularization_stoichiometry} using the two effective reaction currents $\smodcyclecurrel{\emer_\lmb}$ and $\smodcyclecurrel{\emer_\lmb'}$ as coefficients, i.e.,
\begin{equation}
\smodsscurr{\lmb} 
=
\smodcyclecurrel{\emer_\lmb} \kbordermatrix{
&		\cr
& 1		\cr
& 1  		\cr
& 1  		\cr
& 0  		\cr
& 0  		\cr
  }
+ \smodcyclecurrel{\emer_\lmb'} \kbordermatrix{
&		\cr
& 0		\cr
&-1  		\cr
& 0  		\cr
& 1  		\cr
& 1  		\cr
  }
\,,
\end{equation}
which leads to 
{\small\begin{equation}
\begin{split}
\smodcyclecurrel{\emer_\lmb} 
= &\smodsscurrel{\lmb_1} = \smodsscurrel{\lmb_3}\\
= &\frac{\consquantity_{\ch{E}_\lmb}}{\mathcal D_{\lmb}}\Big( k_{+\lmb_{1}}k_{+\lmb_{3}}\big(k_{+\lmb_{2}}( k_{-\lmb_{4}} + k_{+\lmb_{5}}) + k_{+\lmb_{4}}k_{+\lmb_{5}}[\ch{N_{$\lma$}}] \big)[\ch{F}]\\
- & k_{-\lmb_{1}}k_{-\lmb_{3}}\big(k_{-\lmb_{2}}(k_{-\lmb_{4}} + k_{+\lmb_{5}}) + k_{-\lmb_{4}}k_{-\lmb_{5}}[\ch{N_{$\lmb$}}]\big)[\ch{W}] \Big)
\end{split}
\label{eq:model2_effcurr1}
\end{equation}}
and
{\small\begin{equation}
\begin{split}
\smodcyclecurrel{\emer_\lmb'}
= &\smodsscurrel{\lmb_4} = \smodsscurrel{\lmb_5}\\
= & \frac{\consquantity_{\ch{E}_\lmb}}{\mathcal D_{\lmb}}\Big( k_{+\lmb_{4}}k_{+\lmb_{5}} \big(k_{+\lmb_{1}}(k_{-\lmb_{2}} + k_{+\lmb_{3}}) [\ch{F}] + k_{-\lmb_{2}}k_{-\lmb_{3}}  [\ch{W}]\big)[\ch{N_{$\lma$}}] \\
- & k_{-\lmb_{4}}k_{-\lmb_{5}}\big(k_{-\lmb_{3}}(k_{-\lmb_{1}} + k_{+\lmb_{2}})[\ch{W}] + k_{+\lmb_{1}}k_{+\lmb_{2}}[\ch{F}]\big)[\ch{N_{$\lmb$}}]\Big)\,.
\end{split}
\label{eq:model2_effcurr2}
\end{equation}}

In general, the diagrammatic method~\cite{King1956, Hill1966} provides the steady-state concentrations of the internal species of a module and then, by applying mass-action kinetics, its steady-state current vector.

\subsection{Numerical Strategy\label{sub:effcurr_numerical}}

\begin{figure}[t]
  \centering
  \includegraphics[width=0.9999\columnwidth]{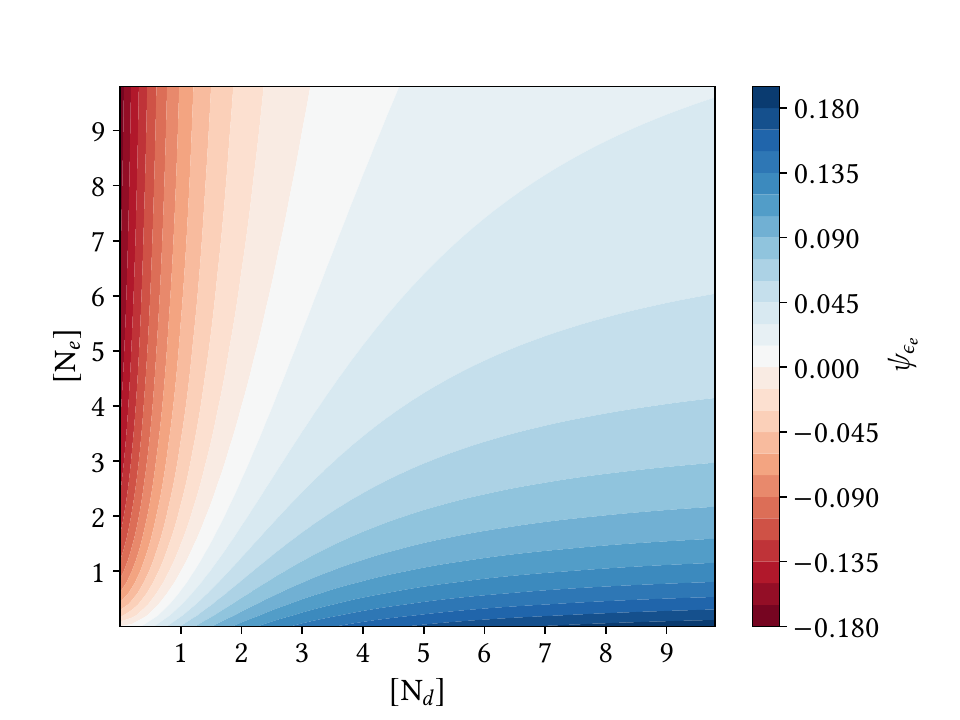}
\caption{Current of the effective reaction~\eqref{eq:module5_eff_rct} for different values of the concentrations of the terminal species $\ch{N}_d$ and $\ch{N}_e$.
We use $1/k_{-\lme_1}$ and $k_{-\lme_1}/k_{+\lme_1}$ as units of measure for time and concentration, respectively.
We assume $k_{+\lme_1} = k_{+\lme_2}= k_{-\lme_3}= k_{-\lme_4}$, $k_{-\lme_1} = k_{-\lme_2}= k_{+\lme_3}= k_{+\lme_4}$, and $[\ch{M}]+[\ch{M^{*}}]+[\ch{A_2}]+[\ch{A^{*}_2}] = k_{-\lme_1}/k_{+\lme_1}$.}
\label{fig:curr_nl}
\end{figure}

When the internal reactions are nonlinear (i.e., not pseudo-first-order reactions), but the kinetic constants of the internal reactions are known, the current-concentration characteristic can be determined numerically. 
We illustrate this procedure for the (purple) module~$\lme$ in Fig.~\ref{fig:modularization}a, where the internal species \ch{M}, \ch{M^{*}}, \ch{A_2} and \ch{A^{*}_2} react via the chemical reactions $\lme_1$, $\lme_2$, $\lme_3$, and $\lme_4$ 
with the terminal species~\ch{N_{$\lmd$}} and~\ch{N_{$\lme$}}.
Reaction $\lme_{2}$ and $\lme_{4}$ are bimolecular reactions in \ch{M^{*}} and \ch{M}, respectively.
When the kinetic constants of the internal reactions are known, one can numerically compute the steady-state concentrations of the internal species for different concentrations of the terminal ones, namely, $(\steady{[\ch{M^{}}]},\steady{[\ch{M^{*}}]},\steady{[\ch{A_2^{}}]},\steady{[\ch{A_2^{*}}]})$ for every value of $([\ch{N_{$\lmd$}}],[\ch{N_{$\lme$}}])$.
To do so, one can either use algorithms that directly determine the fixed point of the rate equation, or simulate the evolution of the internal concentrations until steady state is reached for fixed concentrations of the terminal species.  
Then, one can repeat the steps of App.~\ref{sub:effcurr_analytical}.
First, the steady-state current vector $\smodsscurr{\lme} =(\smodsscurrel{\lme_1},\smodsscurrel{\lme_2},\smodsscurrel{\lme_3},\smodsscurrel{\lme_4})$ is determined for every value of the concentrations $([\ch{N_{$\lmd$}}],[\ch{N_{$\lme$}}])$ using mass-action kinetics and the numerical determined values of $(\steady{[\ch{M^{}}]},\steady{[\ch{M^{*}}]},\steady{[\ch{A_2^{}}]},\steady{[\ch{A_2^{*}}]})$.
Second, $\smodsscurr{\lme}$ is written as a linear combination of cycles.
In this case the stoichiometric matrix
admits one emergent cycle $\cycle_{\emer_\lme}$ (given in Fig.~\ref{fig:modularization_stoichiometry})
whose corresponding effective reaction is specified in~\eqref{eq:module5_eff_rct}.
Hence, $\smodsscurr{\lme} = \smodcyclecurrel{\emer_\lme}\cycle_{\emer_\lme}$, which leads to the effective reaction current
\begin{equation}
\smodcyclecurrel{\emer_\lme}=\smodsscurrel{\lme_1}/2=\smodsscurrel{\lme_2}=\smodsscurrel{\lme_3}=\smodsscurrel{\lme_4}\,,
\end{equation}
shown in Fig.~\ref{fig:curr_nl} for a specific set of kinetic constants $\{k_{\pm\lme_\muteindex}\}_{\muteindex=1,2,3,4}$.


\section{Underlying Assumptions and Limitations of the Circuit Theory\label{app:limits}}

\subsection{Multistability~\label{sub:instability}}
The circuit description given in Sec.~\ref{sec:circuit_description} and Eq.~\eqref{eq:rate_eq_ct}  implicitly assumes that the effective reaction currents are fully determined by the values of the concentrations of the terminal species only.
This imposes that the internal species of the modules relax instantaneously towards a \textit{unique} steady state for all values of the concentrations of the terminal species.
When a module has multiple steady states, its dynamics cannot, in general, be characterized in terms of the concentrations of the terminal species only.

\begin{figure}[t]
  \centering
  \includegraphics[width=0.99\columnwidth]{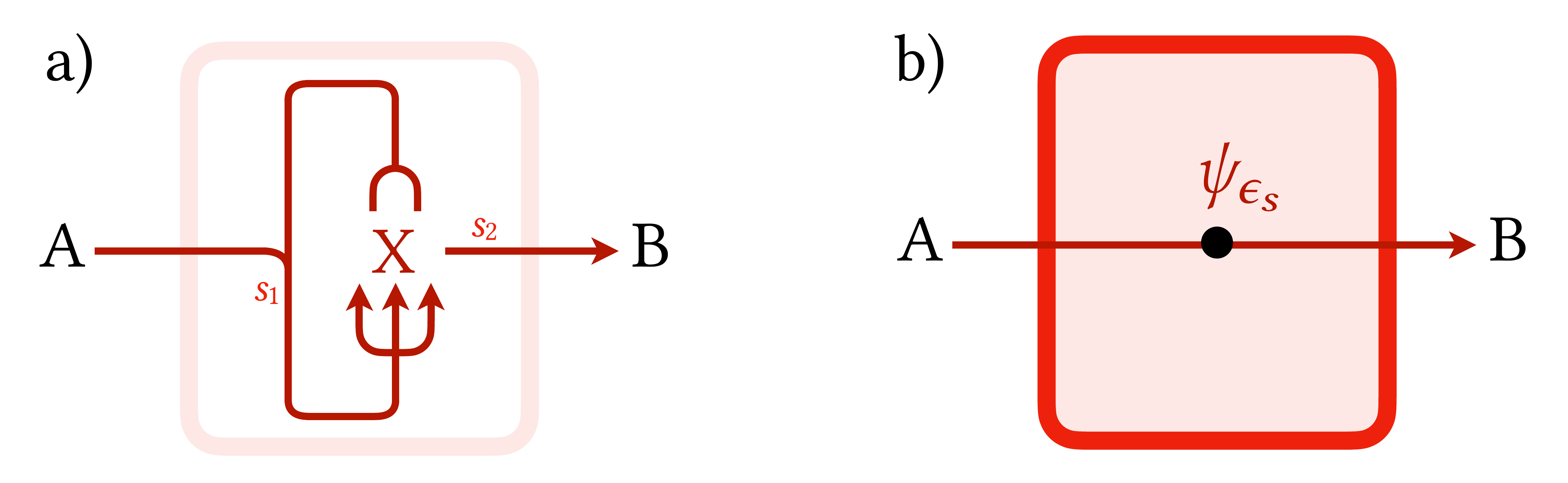}
\caption{Elementary (a) and circuit description (b) of the Schl\"ogl model~\cite{Vellela2009}.}
\label{fig:schlogl}
\end{figure}
Consider for instance the module in Fig~\ref{fig:schlogl} which interconverts the terminal species $\ch{A}$ and $\ch{B}$ through the autocatalytic chemical reactions
\begin{equation}
\begin{split}
\ch{A + 2 X &<=>[ $+\lms_{1}$ ][ $-\lms_{1}$ ] 3 X }\\
\ch{X &<=>[ $+\lms_{2}$ ][ $-\lms_{2}$ ] B }
\end{split}
\label{eq:schlogl_rct}
\end{equation}
with the internal species $\ch{X}$ and a single effective reaction
\begin{equation}
\ch{A <=>[ $\emer_{\lms}$ ][ ] B }\,.
\end{equation}
This is the well known Schl\"ogl model~\cite{Vellela2009}, displaying bistability far from equilibrium when $k_{+\lms_2} \gtrsim 1.7$ (using specific units of measure such that $k_{+\lms_1}[\ch{A}] = k_{-\lms_2}[\ch{B}]=1$). 
\begin{figure}[t]
  \centering
  \includegraphics[width=0.99\columnwidth]{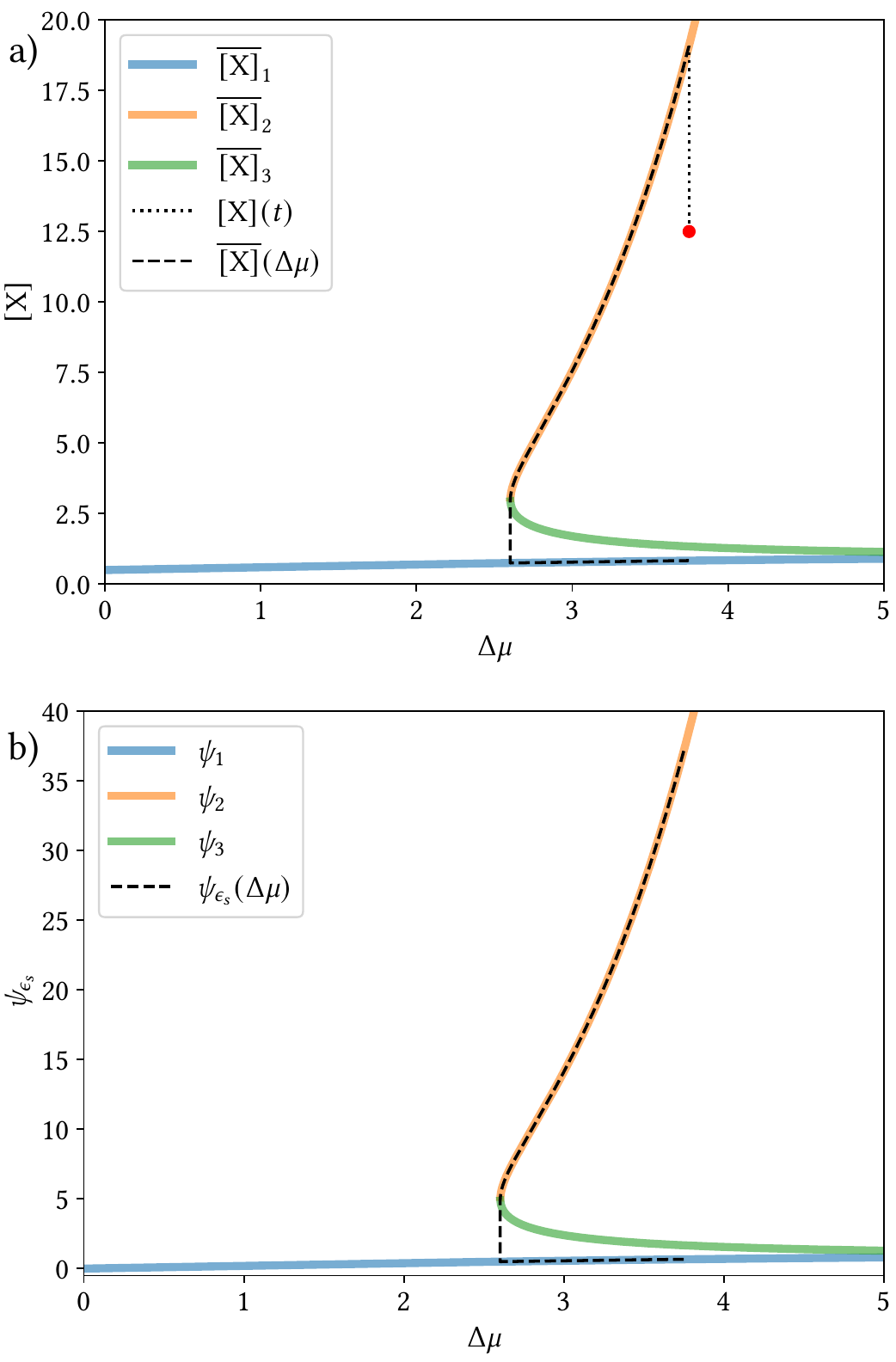}
\caption{Steady-state concentration (a) and effective current (b) of the Schlogl module~\ref{fig:schlogl} for different values of  the chemical potential difference between the terminal species $\ch{A}$ and $\ch{B}$, i.e.,  $\Delta\chempotential = \chempotential_{\ch{A}} - \chempotential_{\ch{B}}$.
The red dot represents an initial concentration $[\ch{X}](0)$ of the module which relaxes along the dotted line towards the corresponding steady state.
The dashed lines specify the value of the steady state concentration $\steady{[\ch{X}]}$ and effective current $\modcyclecurrel_{\emer_{\lms}}$ when 
the value of $\Delta\chempotential$ is decreased from 3.75 to 2.6 and then increased back to 3.75 assuming that $[\ch{X}](0) = 12.5$.
We use units of measure such that $k_{+\lms_1}[\ch{A}] = k_{-\lms_2}[\ch{B}]=1$ which, together with the local detailed balance condition, imposes $ k_{-\lms_1} =  k_{+\lms_2} e^{-\Delta\chempotential/RT}$.
We assume  $k_{+\lms_2}$ = 2.}
\label{fig:schlogl_plots}
\end{figure}

Indeed, when the chemical potential difference $\Delta\chempotential $ between the terminal species $\ch{A}$ and $\ch{B}$ is small enough,
e.g., $\Delta\chempotential <2.6$ if $k_{+\lms_2} = 2$, 
the steady state concentration $\steady{[\ch{X}]}$ of internal species has the unique value $\steady{[\ch{X}]}_1$ represented by the blue line in Fig.~\ref{fig:schlogl_plots}a.
Correspondingly, the effective reaction current $\modcyclecurrel_{\emer_{\lms}}$ has the unique value $\modcyclecurrel_{1} $ represented by the blue line in Fig.~\ref{fig:schlogl_plots}b.
On the other hand, when $\Delta\chempotential > 2.6$ if $k_{+\lms_2} = 2$, there are two possible stable steady state concentrations $\steady{[\ch{X}]}_1$  and $\steady{[\ch{X}]}_2$ represented by the blue and orange line in Fig.~\ref{fig:schlogl_plots}a, respectively, and one unstable steady state concentration $\steady{[\ch{X}]}_3$ represented by the green line in Fig.~\ref{fig:schlogl_plots}a.
Correspondingly, the effective reaction current $\modcyclecurrel_{\emer_{\lms}}$ can have three different values $\modcyclecurrel_{1}$, $\modcyclecurrel_{2}$ and $\modcyclecurrel_{3}$ represented by the blue, orange and green line in Fig.~\ref{fig:schlogl_plots}b, respectively.

This implies that the steady state to which the internal species relaxes and, consequently, the effective reaction current are not uniquely determined by the terminal species.
To see this, immagine that the initial concentration of internal species is $[\ch{X}] (0) = 12.5$ and the concentration of the terminal species are such that $\Delta\chempotential = 3.75$ (red point in Fig.~\ref{fig:schlogl_plots}a).
Then, assuming the time scale separation holds, after an rapid transient (dotted line in Fig.~\ref{fig:schlogl_plots}a), the concentration of internal species reaches the steady state $\steady{[\ch{X}]} = \steady{[\ch{X}]}_2=19$ and the effective current becomes $\modcyclecurrel_{\emer_{\lms}}= \modcyclecurrel_{2}\simeq37$.
If the concentrations of the terminal species change (because of the dynamics of the module and the coupling with other possible modules in a large CRN) in such a way that $\Delta\chempotential$ decreases until $\Delta\chempotential = 2.6$, 
the steady-state concentration $\steady{[\ch{X}]}$ and  effective current $\modcyclecurrel_{\emer_{\lms}}$ will follow the black dashed lines overlapping the orange lines in Fig.~\ref{fig:schlogl_plots}a and Fig.~\ref{fig:schlogl_plots}b, respectively.
Once $\Delta\chempotential < 2.6$, the steady state concentration $\steady{[\ch{X}]}$ jumps from $\steady{[\ch{X}]}_2$ to $\steady{[\ch{X}]}_1$. 
If finally the concentrations of the terminal species change in such a way that $\Delta\chempotential$ is increased back until $\Delta\chempotential = 3.75$, 
the steady-state concentration $\steady{[\ch{X}]}$ and  effective current $\modcyclecurrel_{\emer_{\lms}}$ will follow the black dashed lines overlapping the blue lines in Fig.~\ref{fig:schlogl_plots}a and Fig.~\ref{fig:schlogl_plots}b, respectively.

In general, this kind of evolution of a module cannot be obtained by  a circuit description accounting only for the terminal species since different values of the current correspond to the same values of the concentrations of the terminal species.
Nevertheless, if the current-concentration characteristic resolves the multiple steady states 
(like we have done above for the Schl\"ogl model),
the circuit description still holds.

\subsection{Open CRNs as Modules~\label{sub:open_as_module}}
\begin{figure}[t]
  \centering
\includegraphics[width=0.9999\columnwidth]{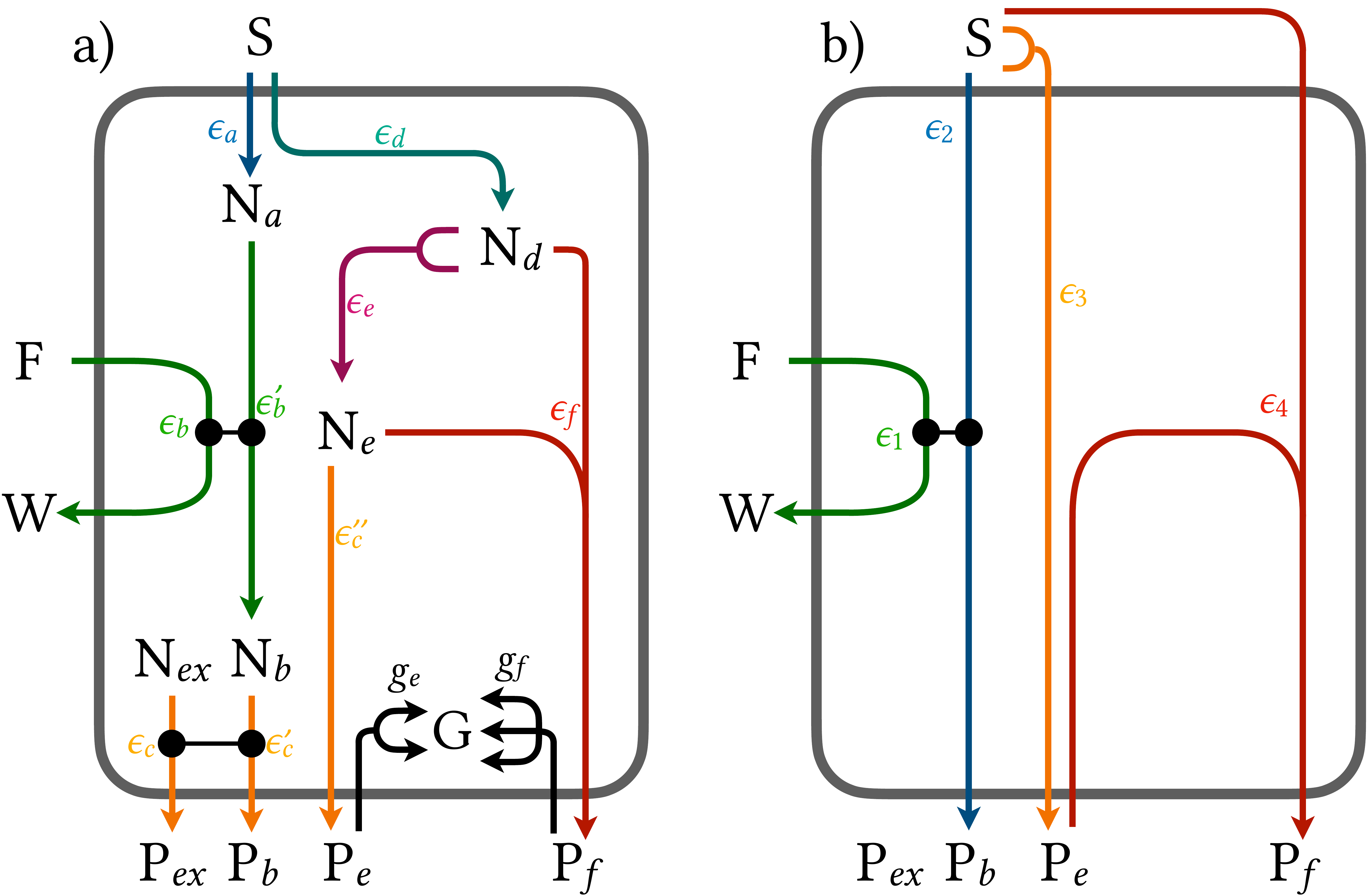}
\caption{
First (a) and second (b) order circuit description of the CRN in Fig.~\ref{fig:modularization}a.
Note that \ch{P_{$ex$}} is not interconverted by the effective reactions~\eqref{eq:eff_rct_final} and thus no arrows connect it to the module in the second-order circuit description.
}
\label{fig:modularization_final_CRN}
\end{figure}

In the circuit description of the open CRN in Fig.~\ref{fig:modularization}b, the species $\ch{S}$, $\ch{F}$, $\ch{W}$, $\ch{P_{$ex$}}$, $\ch{P_{$\lmb$}}$, $\ch{P_{$\lme$}}$, and $\ch{P_{$\lmf$}}$ are exchanged with the environment. 
Let us assume now that the environment is constituted by other chemical processes.
In this case, $\ch{S}$, $\ch{F}$, $\ch{W}$, $\ch{P_{$ex$}}$, $\ch{P_{$\lmb$}}$, $\ch{P_{$\lme$}}$, and $\ch{P_{$\lmf$}}$ are involved in the chemical reactions of both the CRN in Fig.~\ref{fig:modularization}b and the environment.
Namely, they play the role of terminal species coupling the CRN to the environment, and hence the CRN in Fig.~\ref{fig:modularization}b can be treated as a \textit{module} like in Fig.~\ref{fig:modularization_final_CRN}a.

As done for the modules in Fig.~\ref{fig:modularization}a, also the module in Fig.~\ref{fig:modularization_final_CRN}a can be further coarse grained into what we could call a second-order circuit description given in Fig.~\ref{fig:modularization_final_CRN}b (assuming that the time scale separation between internal and terminal species holds).
To do so, we follow the same strategy as before. 
First, we determine the effective reactions by looking for the emergent cycles of the stoichiometric matrix in Eq.~\eqref{eq:stoichiometric_circuit},
where black horizontal line now splits $\matSeff$ into the substoichiometric matrix $\matSeffQ$ for the internal species (i.e., $\ch{N_{$\lma$}}$, $\ch{N_{$\lmb$}}$, $\ch{N_{$ex$}}$, $\ch{N_{$\lmd$}}$, $\ch{N_{$\lme$}}$, and $\ch{G}$) and the substoichiometric matrix $\matSeffP$ for the terminal species (i.e., $\ch{S}$, $\ch{F}$, $\ch{W}$, $\ch{P_{$ex$}}$, $\ch{P_{$\lmb$}}$, $\ch{P_{$\lme$}}$, and $\ch{P_{$\lmf$}}$).
The right-null vectors of $\matSeffQ$ include the internal cycle~\eqref{eq:internal_CD} and the emergent cycles
\begin{subequations}
\begin{equation}
\cyclect_{\emer_1}=
 \kbordermatrix{
   						 &\cr
    \color{g}{\emer_{\lma}}    	& 0		\cr
    \color{g}{\emer_{\lmb}} 		& 1  		\cr
    \color{g}{\emer_{\lmb}'}  	& 0  		\cr
    \color{g}{\emer_{\lmc}}	  	& 0  		\cr
    \color{g}{\emer_{\lmc}'}	  	& 0  		\cr
    \color{g}{\emer_{\lmc}''}	  	& 0  		\cr
    \color{g}{\emer_{\lmd}}	  	& 0  		\cr
    \color{g}{\emer_{\lme}}	  	& 0  		\cr
    \color{g}{\emer_{\lmf}}	  	& 0  		\cr
    \color{g}{{\lmge}}		  	& 0  		\cr
    \color{g} {{\lmgf}}		  	& 0  		\cr
  }\,,\text{ }\text{ }\text{ }\text{ }\text{ }\text{ }\text{ }\text{ }
\cyclect_{\emer_2}=
 \kbordermatrix{
   						 &\cr
    \color{g}{\emer_{\lma}}    	& 1		\cr
    \color{g}{\emer_{\lmb}} 		& 0  		\cr
    \color{g}{\emer_{\lmb}'}  	& 1  		\cr
    \color{g}{\emer_{\lmc}}	  	& 0  		\cr
    \color{g}{\emer_{\lmc}'}	  	& 1  		\cr
    \color{g}{\emer_{\lmc}''}	  	& 0  		\cr
    \color{g}{\emer_{\lmd}}	  	& 0  		\cr
    \color{g}{\emer_{\lme}}	  	& 0  		\cr
    \color{g}{\emer_{\lmf}}	  	& 0  		\cr
    \color{g}{{\lmge}}		  	& 0  		\cr
    \color{g} {{\lmgf}}		  	& 0  		\cr
  }\,,
\end{equation}
\begin{equation}
\cyclect_{\emer_3}=
 \kbordermatrix{
   						 &\cr
    \color{g}{\emer_{\lma}}    	& 0		\cr
    \color{g}{\emer_{\lmb}} 		& 0  		\cr
    \color{g}{\emer_{\lmb}'}  	& 0  		\cr
    \color{g}{\emer_{\lmc}}	  	& 0  		\cr
    \color{g}{\emer_{\lmc}'}	  	& 0  		\cr
    \color{g}{\emer_{\lmc}''}	  	& 1  		\cr
    \color{g}{\emer_{\lmd}}	  	& 2  		\cr
    \color{g}{\emer_{\lme}}	  	& 1  		\cr
    \color{g}{\emer_{\lmf}}	  	& 0  		\cr
    \color{g}{{\lmge}}		  	& 0  		\cr
    \color{g} {{\lmgf}}		  	& 0  		\cr
  }\,,\text{ }\text{ }\text{ }\text{ }\text{ }\text{ }\text{ }\text{ }
\cyclect_{\emer_4}=
 \kbordermatrix{
   						 &\cr
    \color{g}{\emer_{\lma}}    	& 0		\cr
    \color{g}{\emer_{\lmb}} 		& 0  		\cr
    \color{g}{\emer_{\lmb}'}  	& 0  		\cr
    \color{g}{\emer_{\lmc}}	  	& 0  		\cr
    \color{g}{\emer_{\lmc}'}	  	& 0  		\cr
    \color{g}{\emer_{\lmc}''}	  	&-1  		\cr
    \color{g}{\emer_{\lmd}}	  	& 1  		\cr
    \color{g}{\emer_{\lme}}	  	& 0  		\cr
    \color{g}{\emer_{\lmf}}	  	& 1  		\cr
    \color{g}{{\lmge}}		  	& 0  		\cr
    \color{g} {{\lmgf}}		  	& 0  		\cr
  }\,,
\end{equation}
\end{subequations}
which correspond to the following effective reactions between the terminal species:
\begin{equation}
\begin{split}
\ch{F &<=>[ $\emer_1$ ][ $$ ] W } \,,\\
\ch{S &<=>[ $\emer_2$ ][ $$ ] P_{$\lmb$} }\,,\\
\ch{2 S &<=>[ $\emer_3$ ][ $$ ] P_{$\lme$} }\,,\\
\ch{S + P_{$\lme$} &<=>[ $\emer_4$ ][ $$ ] P_{$\lmf$} }\,.
\end{split}
\label{eq:eff_rct_final}
\end{equation}
Note that the current-concentration characteristic of these reactions cannot be determined using the diagrammatic method~\cite{King1956, Hill1966} as the dynamics of the module in Fig.~\ref{fig:modularization_final_CRN}a (given in Eq.~\eqref{eq:rate_eq_ct}) does not follow mass-action kinetics, and one should therefore rely on the numerical~\ref{sub:effcurr_numerical} or the experimental strategy (see Sec.~\ref{sec:currents}). 

\subsection{Further Decomposition of the Modules\label{sub:decomposition}}
\begin{figure}[t]
  \centering
  \includegraphics[width=0.99\columnwidth]{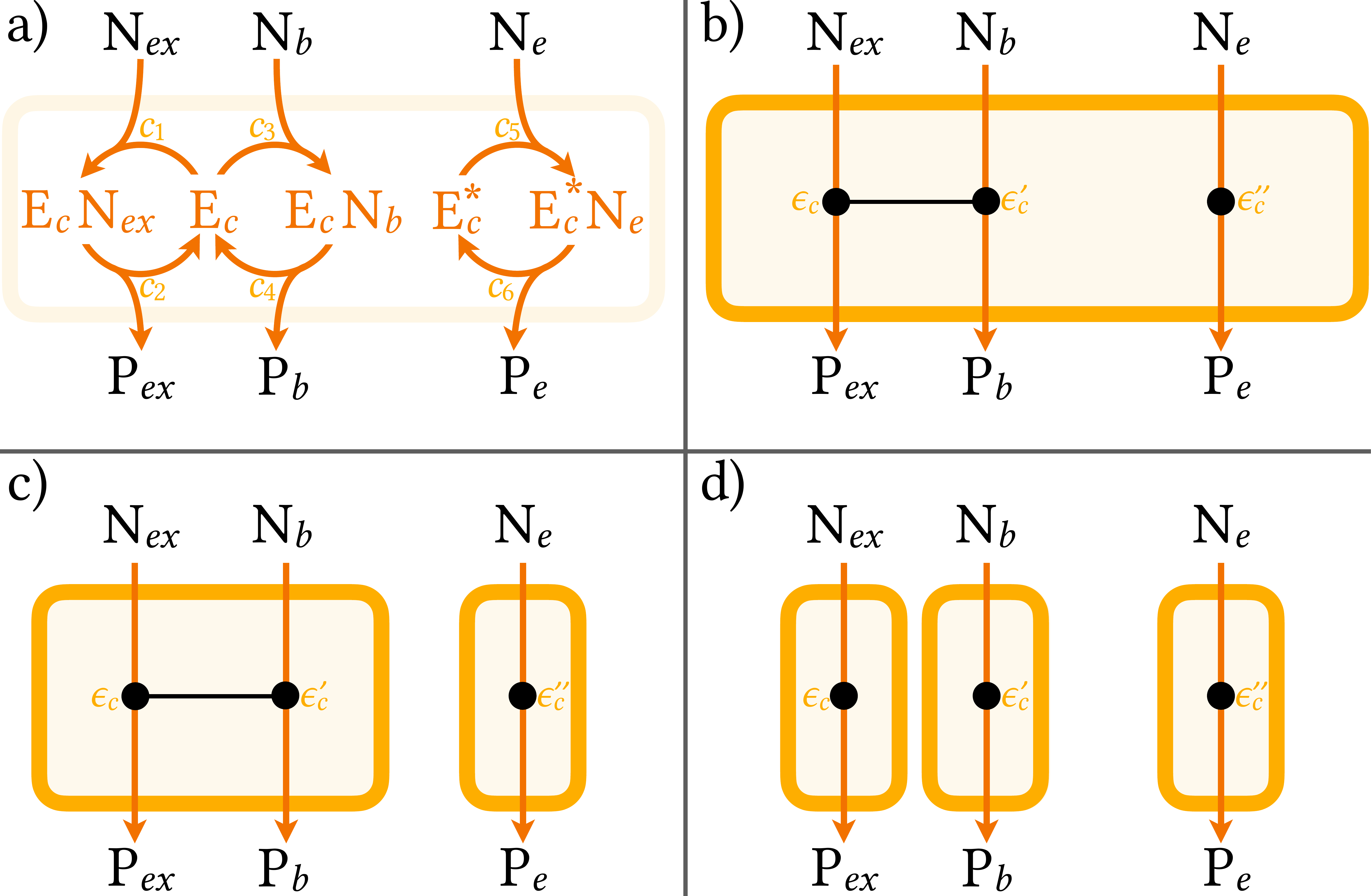}
\caption{Elementary (a) and circuit description (b) of the (orange) module~$\lmc$ in Fig.~\ref{fig:modularization} together with a consistent (c) and inconsistent (d) decomposition in submodules.}
\label{fig:futher_modularization}
\end{figure}
In the circuit description, each module is coarse grained into at least one effective reaction between terminal species.
When a module (like the (orange) module~$\lmc$ in Fig.~\ref{fig:modularization}a and~\ref{fig:futher_modularization}a) has more than one effective reaction (given in Fig.~\ref{fig:futher_modularization}b), one can ask if it can be split into independent (sub)modules corresponding to a single effective reaction each (like in Fig.~\ref{fig:futher_modularization}d).
This can be done only if each (sub)module has a unique set of internal species. Indeed, the procedure to split a module into submodules is exactly the same as the one to split a generic CRN into modules and must satisfy the same assumptions. 

One can also determine if a module can be split into submodules by examining the currents of the effective reactions without analyzing the internal species.
When the currents of some effective reactions depend only on a subset of terminal species, they involve a unique set of internal species and constitute, by definition, an independent (sub)module.
To see this, we consider the effective reaction currents (derived using the diagrammatic method~\cite{King1956, Hill1966}) of the (orange) module~$\lmc$ in Fig.~\ref{fig:modularization}b:
\begin{subequations}\small{
\begin{align}
\modcyclecurrel_{\emer_{\lmc}} &= \frac{\consquantity_{\ch{E}_\lmc}}{\mathcal D_{\lmc}} (k_{-\lmc_{3}} + k_{+\lmc_{4}}) \left( k_{+\lmc_{1}}k_{+\lmc_{2}}[\ch{N_{$ex$}}] - k_{-\lmc_{1}}k_{-\lmc_{2}}[\ch{P_{$ex$}} ]\right)\,,\\
\modcyclecurrel_{\emer_{\lmc}'} &= \frac{\consquantity_{\ch{E}_\lmc}}{\mathcal D_{\lmc}} (k_{-\lmc_{1}} + k_{+\lmc_{2}}) \left( k_{+\lmc_{3}}k_{+\lmc_{4}}[\ch{N_{$\lmb$}}] - k_{-\lmc_{3}}k_{-\lmc_{4}}[\ch{P_{$\lmb$}} ]\right)\,,\\
\modcyclecurrel_{\emer_{\lmc}''} &= \frac{\consquantity_{\ch{E^{*}_{$\lmc$}}}}{\mathcal D_{\lmc}^{*}} \left( k_{+\lmc_{5}}k_{+\lmc_{6}}[\ch{N_{$\lme$}}] - k_{-\lmc_{5}}k_{-\lmc_{6}}[\ch{P_{$\lme$}} ]\right)\,,
\end{align}}
\end{subequations}
where $\mathcal D_{\lmc}$ and $\mathcal D_{\lmc}^{*} $ are given by
\begin{subequations} 
\begin{equation}
\begin{split}
\mathcal D_{\lmc} = 
&(k_{-\lmc_{1}} + k_{+\lmc_{2}}) (k_{-\lmc_{3}} + k_{+\lmc_{4}})\\
& + (k_{-\lmc_{3}} + k_{+\lmc_{4}}) (k_{+\lmc_{1}}[\ch{N_{$ex$}}] + k_{-\lmc_{2}}[\ch{P_{$ex$}}])\\
& + (k_{-\lmc_{1}} + k_{+\lmc_{2}}) (k_{+\lmc_{3}}[\ch{N_{$\lmb$}}] + k_{-\lmc_{4}}[\ch{P_{$\lmb$}}])
\end{split}
\end{equation}
\begin{equation}
\mathcal D_{\lmc}^{*} = k_{+\lmc_{5}} [\ch{N_{$\lme$}}] + k_{-\lmc_{5}} + k_{+\lmc_{6}} + k_{-\lmc_{6}} [\ch{P_{$\lme$}}]
\end{equation}
\end{subequations}
respectively.
On the one hand, the reaction current $\modcyclecurrel_{\emer_{\lmc}''}$ depends only on the concentration of the terminal species \ch{N_{$\lme$}} and \ch{P_{$\lme$}}.
On the other hand, the reaction current $\modcyclecurrel_{\emer_{\lmc}}$ (resp. $\modcyclecurrel_{\emer_{\lmc}'}$) does not only depend on the concentration of \ch{N_{$ex$}} and \ch{P_{$ex$}} (resp. \ch{N_{$\lmb$}} and \ch{P_{$\lmb$}}), but also on the concentration of \ch{N_{$\lmb$}} and \ch{P_{$\lmb$}} (resp. \ch{N_{$ex$}} and \ch{P_{$ex$}}) via $\mathcal D_{\lmc}$.
Thus, only reaction $\emer_{\lmc}''$ can be treated as an independent module (as represented in Fig.~\ref{fig:futher_modularization}c). 
The coupling between reaction $\emer_{\lmc}$ and $\emer_{\lmc}'$ is a direct consequence of sharing the internal species \ch{E_{$\lmc$}}:
whether or not this species is available for one effective reaction depends on how much is involved in the other.
The two reactions $\emer_{\lmc}$ and $\emer_{\lmc}'$ must therefore be considered as part of the same module despite closely resembling a Michaelis and Menten mechanism.

\subsection{Experimental Derivation of the Effective Reactions\label{sub:effrct_via_experiment}}
The effective reactions of the modules have been identified by deriving the emergent cycles from the stoichiometry of the elementary reactions in Sec.~\ref{sec:module_def} and App.~\ref{app:def_chem_module_eff}.
If the elementary stoichiometry is not known, the effective reactions can still be determined by using a similar approach as the one implemented in Sec.~\ref{sec:currents}.

Consider for instance to transfer the (blue) module~$\lma$ in Fig.~\ref{fig:modularization} in a reactor where the concentration of the terminal species \ch{S} and \ch{N_{$\lma$}} can be maintained constant via the exchange currents $\ssexcurrel_{\ch{S}}$ and $\ssexcurrel_{\ch{N_{$\lma$}}}$ as done in Fig.~\ref{fig:IVcharacteristicA}.
By measuring these exchange currents, one would observe that they always satisfy
\begin{equation}
\ssexcurrel_{\ch{S}}  = - \ssexcurrel_{\ch{N_{$\lma$}}}\label{eq:der_sotiochiometry_1}\,.
\end{equation}
Since these exchange currents, as already pointed out, balance the variations of the concentrations due to the effective reaction,
Eq.~\eqref{eq:der_sotiochiometry_1} means that every time the current $\ssexcurrel_{\ch{S}}$ provides (resp. extracts) 1 molecule of \ch{S} because consumed (resp. produced) by the effective reaction, the current $\ssexcurrel_{\ch{N_{$\lma$}}}$ extracts (resp. provides) 1 molecule of \ch{N_{$\lma$}}.
This implies that the net stoichiometry of the effective reaction must be the one represented in~\eqref{eq:module1_eff_rct}.

When the same approach is applied to the (purple) module~$\lme$ in Fig.~\ref{fig:modularization}a, one would observe that 
\begin{equation}
\ssexcurrel_{\ch{N_{$\lmd$}}}/2  = - \ssexcurrel_{\ch{N_{$\lme$}}}\label{eq:der_sotiochiometry_2}\,,
\end{equation}
which physically means that every time the current $\ssexcurrel_{\ch{N_{$\lmd$}}}$ provides (resp. extracts) 2 molecules of \ch{N_{$\lmd$}} because consumed (resp. produced) by the effective reaction, the current $\ssexcurrel_{\ch{N_{$\lme$}}}$ extracts (resp. provides) 1 molecule of \ch{N_{$\lme$}}.
Thus, the net stoichiometry of the effective reaction must be the one represented in~\eqref{eq:module5_eff_rct}.

\begin{figure}[t]
  \centering
  \includegraphics[width=0.99\columnwidth]{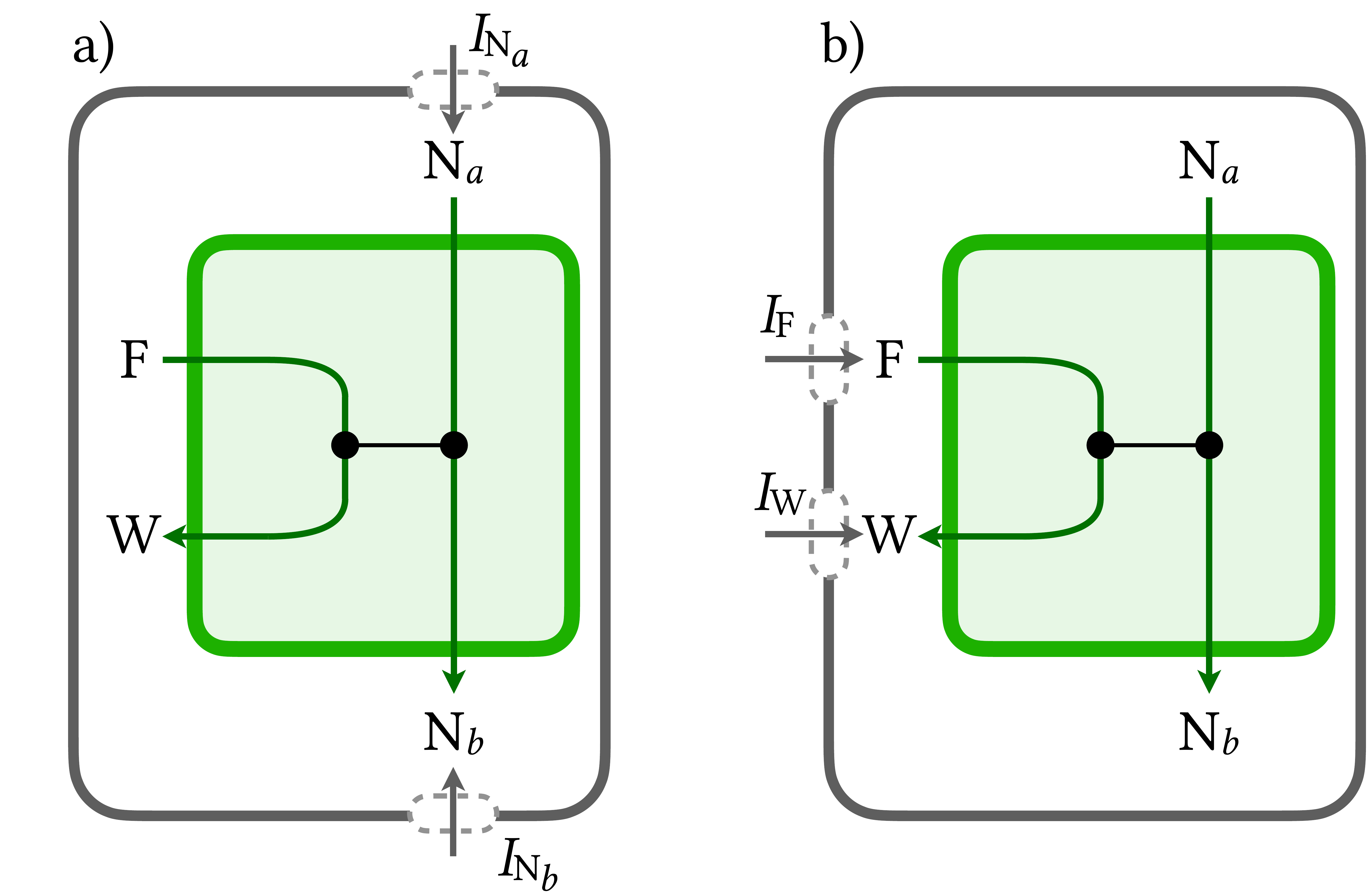}
\caption{Effective reactions of the (green) module~$b$ in  Fig.~\ref{fig:modularization} in a reactor, similar to the membrane reactor used in Ref.~\cite{Sorrenti2017}, where either the concentrations of \ch{N_{$\lma$}}, \ch{N_{$\lmb$}} (a) or \ch{F} and \ch{W} (b) are controlled by exchange processes whose currents are specified by  $ \ssexcurrel_{\ch{N_{$\lma$}}}$, $\ssexcurrel_{\ch{N_{$\lmb$}}}$, $\ssexcurrel_{\ch{F}}$, and $\ssexcurrel_{\ch{W}}$.}
\label{fig:stoichiometryB}
\end{figure}
Modules with more than two terminal species may have many effective reactions.
This complicates determining their stoichiometry since the exchange currents balance the variations of the concentrations due to \textit{all} the effective reactions. 
To recognize the contribution of each effective reaction, we proceed as follows for the (green) module $\lmb$ in Fig.~\ref{fig:modularization}.
We group the terminal species in all possible combinations, which means that we consider 
all two species combinations, i.e., $(\ch{N_{$\lma$}},\ch{N_{$\lmb$}})$, $(\ch{N_{$\lma$}},\ch{F})$, $(\ch{N_{$\lma$}},\ch{W})$, $(\ch{N_{$\lmb$}},\ch{F})$, $(\ch{N_{$\lmb$}},\ch{W})$, and  $(\ch{F},\ch{W})$, 
all three species combinations, i.e., $(\ch{N_{$\lma$}},\ch{N_{$\lmb$}}, \ch{F})$, $(\ch{N_{$\lma$}},\ch{N_{$\lmb$}}, \ch{W})$, $(\ch{N_{$\lma$}},\ch{F}, \ch{W})$, $(\ch{N_{$\lmb$}},\ch{F}, \ch{W})$,
and the only four species combination, i.e., $(\ch{N_{$\lma$}},\ch{N_{$\lmb$}}, \ch{F},  \ch{W})$.
We then maintain the concentrations of only the terminal species belonging to a specific combination constant.
The concentrations of the other terminal species are free to evolve as they were internal species of the module. 
By measuring the exchange currents for every combination, we can determine whether the corresponding species are involved in an effective reaction or not and the corresponding stoichiometry.
For the combination $(\ch{N_{$a$}},\ch{N_{$b$}})$ (illustrated in Fig.~\ref{fig:stoichiometryB}a), the exchange currents always satisfy 
\begin{equation}
\ssexcurrel_{\ch{N_{$\lma$}}} = -\ssexcurrel_{\ch{N_{$\lmb$}}}\,,
\end{equation}
after a transient dynamics due to the relaxation of the not-controlled terminal species to steady state.
As for the  (blue) module $\lma$, this means that every time the current $\ssexcurrel_{\ch{N_{$\lma$}}}$ provides (resp. extracts) 1 molecule of \ch{N_{$\lma$}} because consumed (resp. produced) by an effective reaction, the current $\ssexcurrel_{\ch{N_{$\lmb$}}}$ extracts (resp. provides) 1 molecule of \ch{N_{$\lmb$}}. 
Thus, there is an effective reaction interconverting \ch{N_{$\lma$}} into \ch{N_{$\lmb$}} with the stoichiometry specified in~\eqref{eq:module2_eff_rct2}.
By repeating the same operation for all the other two species combinations, we would find that only for the combination $(\ch{F},\ch{W})$ (illustrated in Fig.~\ref{fig:stoichiometryB}b), the exchange currents $\ssexcurrel_{\ch{F}}$ and $\ssexcurrel_{\ch{W}}$ do not vanish (after a transient dynamics due to relaxation of the not-controlled terminal species to steady state) and satisfy
\begin{equation}
\ssexcurrel_{\ch{F}} = -\ssexcurrel_{\ch{W}}\,.
\end{equation}
Thus, terminal species \ch{F} and \ch{W} are involved in an effective reaction whose stoichiometry, for the same reasons we already discussed, is specified in ~\eqref{eq:module2_eff_rct1}.
All the other two species combinations are not coupled by an effective reaction.
For the (green) module $\lmb$ in Fig.~\ref{fig:modularization}, we have now determined all the effective reactions since the two identified reactions involve all the terminal species. 
If this was not the case, we should have proceeded by analyzing in the same way all the other combinations until we identified a set of effective reactions involving all terminal species.
For instance, for the (red) module~$\lmf$ in Fig.~\ref{fig:modularization}, the only combination of terminal species leading to non vanishing exchange currents is $(\ch{N_{$\lmd$}},\ch{N_{$\lme$}}, \ch{P_{$\lmf$}})$.
The exchange currents $\ssexcurrel_{\ch{N_{$\lmd$}}}$, $\ssexcurrel_{\ch{N_{$\lme$}}}$, and $\ssexcurrel_{\ch{P_{$\lmf$}}}$ always satisfy
\begin{equation}
\ssexcurrel_{\ch{N_{$\lmd$}}} = \ssexcurrel_{\ch{N_{$\lme$}}} = -\ssexcurrel_{\ch{P_{$\lmf$}}}\,,
\end{equation}
which is consistent with the stoichiometry of the effective reaction given in Eq.~\eqref{eq:module6_eff_rct}.

\bibliography{biblio.bib}
\end{document}